\documentclass[journal]{IEEEtran}
\usepackage{amsmath,amssymb,amsfonts}
\usepackage{algorithm}
\usepackage{algorithmic}
\usepackage[hiresbb]{graphicx}
\usepackage{cite}
\usepackage{bm}
\usepackage{booktabs}
\usepackage{array}
\usepackage{multirow}
\usepackage{makecell}
\usepackage[bookmarks=false,hidelinks]{hyperref}
\hypersetup{
  pdftitle={Sensing-Aided Vehicular Beam Tracking: System Gain Over Link Feedback and Probe-Verified Sensing Invocation},
  pdfauthor={Jeongwan Kang, Miguel Bordallo Lopez, Nhan Thanh Nguyen, Junhyeong Kim, and Sunwoo Kim}
}
\usepackage{pifont}
\newtheorem{remark}{Remark}
\DeclareMathOperator*{\argmax}{arg\,max}

\begin{document}
\bstctlcite{IEEEexample:BSTcontrol}

\title{Sensing-Aided Vehicular Beam Tracking: System Gain Over Link Feedback and Probe-Verified Sensing Invocation}

\author{Jeongwan Kang,~\IEEEmembership{Member,~IEEE}, Miguel Bordallo L\'{o}pez,~\IEEEmembership{Member,~IEEE}, \\ Nhan Thanh Nguyen,~\IEEEmembership{Senior Member,~IEEE}, Junhyeong Kim, and Sunwoo Kim,~\IEEEmembership{Senior Member,~IEEE}
\thanks{This work was supported by the Institute of Information \& Communications Technology Planning \& Evaluation (IITP) grant funded by the Korea government (MSIT) (No. RS-2024-00444230, Development of wireless technology for integrated sensing and communication), and supported in part by Business Finland, Nokia, Keysight and Finwe via the 6G-WISECOM project.
 J. Kang and S. Kim are with the Department of Electronic Engineering, Hanyang University, Seoul, Republic of Korea (e-mail: \{rkdwjddhks77, remero\}@hanyang.ac.kr). M. Bordallo L\'{o}pez is with Computer Science and Engineering, and N. T. Nguyen is with the Centre for Wireless Communications, University of Oulu, Oulu, Finland (e-mail: \{miguel.bordallo, nhan.nguyen\}@oulu.fi). J. Kim is with the 6G Wireless Access System Research Section, Electronics and Telecommunications Research Institute (ETRI), Daejeon, Republic of Korea. He is also with the Major in Information and Communication Engineering, ETRI School, University of Science and Technology (UST), Daejeon, Republic of Korea (e-mail: jhkim41jf@etri.re.kr). (Corresponding author: Sunwoo Kim.)}}

\maketitle

\begin{abstract}
Sensor-based beam prediction is usually evaluated against exhaustive beam training, leaving
open when sensing adds useful information beyond a deployed tracker's link feedback and when
it should be withheld. We separate system gain from the marginal value of sensing in
action-restricted causal replay of recorded $64$-beam sweeps from four vehicular
DeepSense~6G scenarios. The replay exposes only serving-beam and explicitly probed
candidate powers and counts probes and model invocations separately. The proposed
controller combines a scheduled post-setup re-sweep with probe-verified sensing admission
and blacklist backoff. In three scenarios where feedback-only tracking is within
$0.25$\,dB of the recorded-sweep oracle, the controller stays close to feedback-only in
mean loss while rarely invoking the model, whereas always-on sensing performs worse.
In the sole high-innovation scenario, the controller's system gain is $0.82$\,dB
($95\%$ confidence interval $[0.30,1.40]$) over the plain feedback-only tracker and $0.15$\,dB over a probe-limited re-sweeping reference, with fewer
probes than either. Sensing and the re-sweep both correct setup lock loss. At the native decision interval,
enabling sensing in the same controller yields no statistically resolved loss
reduction but cuts probing by $21\%$. In this scenario, accurate sensing reduces loss relative to the sensor-disabled
controller under emulated link noise, staleness or longer decision intervals. Weak or
delayed sensing can instead increase loss despite probe verification. These exploratory
results characterize sensing as a conditional substitute for RF probing or a
source of loss reduction, depending on the limitations of link feedback and on the
accuracy and timeliness of sensing.
\end{abstract}


\section{Introduction}\label{sec:intro}
Millimeter-wave vehicular links require narrow beams to follow a moving
vehicle, while the exhaustive sweeps used to recover alignment consume radio
resources~\cite{heath2016overview, niu2015survey}. The cost is a property of narrow
beams rather than of one band: the larger arrays considered for the $7$--$24$\,GHz
range bring the same alignment problem~\cite{3gpp38820, kang2024uppermidband}. The recorded
$60$-GHz sweeps used here are one instance in which this cost can be measured. Cameras, radar, LiDAR, or position
can instead map the scene to a codebook beam~\cite{demirhan2022radar,ma2025kd,
charan2022challenge}. Such predictors, however, are usually compared with exhaustive
or hierarchical training rather than with a feedback-only tracker driven by the link's
own serving-beam and probe measurements. That comparison measures predictor
\emph{accuracy}, but not what sensing adds beyond the feedback a deployed link already
has. We therefore ask when sensing adds useful information beyond a well-instrumented feedback loop, and when it should be withheld. That loop is a feedback-only tracker given standard refinements such as a re-sweep after link setup.

\begin{figure}[t]
\centering
\includegraphics[width=\columnwidth]{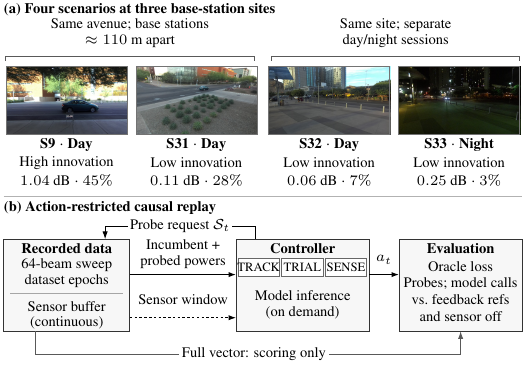}
\caption{Scenarios and causal replay. (a) Representative camera frames from test
passes of the four scenarios~\cite{10144504}; high/low
innovation is the feedback tracker's statistic, and the day/night pair is two
sessions, not a controlled illumination change. The third line gives the two conditions
on which the value of sensing depends: the feedback-only tracker's loss (the room left for sensing) and the deployed vision arm's current-slot top-1 accuracy; only Scenario~9
combines both. (b) After the link-setup sweep, the
policy sees only the incumbent power and the candidates it probes; the full vector
only scores the committed beam $a_t$, and losses are compared with feedback-only
references and with the same controller with its sensor disabled. One epoch is one dataset sample ($111$\,ms in Scenario~9, about $92$\,ms elsewhere).}
\label{fig:scenario_replay}
\end{figure}

The question follows our companion study of long-term radar-only
prediction on the same measurements~\cite{precursor}. In the companion study, all predictions over the horizon come from a single sensor window, so their errors are correlated. Observing the scene again partway through could reduce these errors but costs additional sensing and inference, raising the question of when re-observation is worth its cost.
To price it, we restrict each policy to acquired link measurements and
account separately for candidate-beam probes and model invocations
(Fig.~\ref{fig:scenario_replay}). This restriction matters because open-loop evaluation can be optimistic. In a camera-primed beam-management system, the outage of the camera-only beam estimate at a
signal-to-noise ratio (SNR) threshold rises from $4.7\%$ offline to $33.1\%$ online ($3.6\%$
after refinement with link measurements)~\cite{vibe2026}. Our replay isolates a second source of optimism, oracle re-anchoring (Sec.~\ref{sec:replay_flaw}); all headline results in this paper use the causal protocol.

The policy under test is a controller that decides when the sensing model is
invoked and whether its proposal is used. It opens a probe-verified trial when a sensor window is available, no blacklist is active, and tracker innovation (the disagreement between predicted and probed beam positions) exceeds a threshold, as it usually does after link setup. A trial probes the proposed beam and its two neighbors; sensing is admitted
only when a probed candidate exceeds the serving beam by a fixed margin, and a failed
trial blacklists the sensor. Three epochs after link setup the controller also runs one
coarse re-sweep, the refinement that follows initial access
(Sec.~\ref{sec:voi_controller}).
Adaptive tracker cadence, budget- and age-driven sensing schedules, and
sensing-prior probing with a measured lock margin all have
precedent~\cite{atiekf2024,zakeri2025drl,zakeri2025aoi,probing2026adaptive}. Our
contribution combines probe-verified admission, blacklist backoff and a post-setup
re-sweep, and measures what sensing adds against feedback-only references given the same
re-sweep.

The answer depends on two conditions: how much link feedback leaves to gain, in loss or in
probing effort, and how accurate and timely the sensing information is. In three of four scenarios from three
base-station sites, vehicles pass tens of meters from the base station, the best beam drifts
slowly, and a feedback-only tracker comes within $0.25$\,dB of the recorded-sweep oracle.
There, always-on sensing is worse than feedback-only. In the remaining scenario, vehicles
pass about $11$\,m away, the best beam moves several times faster, and a feedback-only
tracker loses about $1$\,dB, mostly in passes with a weak link-setup sweep. There an accurate camera-based beam predictor (the
vision arm) substitutes external information for candidate-beam measurements at the native
interval and directly reduces loss once feedback is stressed by noise or longer decision
intervals (Secs.~\ref{sec:voi_day}, \ref{sec:voi_decim} and~\ref{sec:voi_tail}). Weak or
late sensing, by contrast, can be harmful, and probe verification does not fully prevent
it.

The novelty of this paper is not a new sensing model but the feedback-referenced
evaluation, the incremental value of sensing it measures, and a probe-verified controller (Table~\ref{tab:positioning}). The main contributions are as follows.
\begin{itemize}
\item \textbf{Causal, feedback-referenced decomposition.} An action-restricted replay of
recorded sweeps charges every candidate-beam probe, counts model invocations separately,
and reveals only acquired observations.
Within it we separate the controller's system gain over feedback-only references, the
marginal value of sensing relative to the same controller with its sensor disabled, and the probe saving. What sensing adds is thus not confused with what the controller's
other components add (Secs.~\ref{sec:model} and~\ref{sec:voi_day}).

\item \textbf{Probe-verified invocation with blacklist backoff.} Measured link power, not
a need estimate or the sensor's own confidence, decides whether a proposal is admitted,
and a failed trial suspends the sensor. The trial enforces one necessary condition, an
immediate measured improvement over the serving beam. With the blacklist and periodic
reverification it protects the link where the sensor is weak at the native interval. One
constant set serves all four scenarios (Secs.~\ref{sec:voi_controller} and~\ref{sec:voi_attrib}).

\item \textbf{Conditional value of sensing.} With an accurate sensor, sensing saves
probes at the native interval and reduces loss once feedback is stressed; weak sensing
under that stress, or a sensor that arrives late, can yield a negative marginal value. What
sensing contributes depends jointly on the limitations of link feedback, the loss it
leaves or the probing it needs, and on the accuracy and timeliness of the sensing
information (Secs.~\ref{sec:voi_day}, \ref{sec:voi_decim} and~\ref{sec:voi_tail}).

\item \textbf{Robustness and comparative evidence.} Candidate controls show that, under the same probing and acquisition rules, a non-informative proposal raises loss. Constants selected without the high-innovation site and a frontier of
tested feedback-only policies corroborate the system gain, and competing sensing policies selected like the controller add
comparative context (Secs.~\ref{sec:voi_controller}, \ref{sec:voi_day} and~\ref{sec:voi_attrib}).
\end{itemize}

\section{Related Work}\label{sec:related}
\subsection{Always-On Sensing-Aided Beam Prediction}
A substantial literature maps sensor observations to codebook beams, for the current
slot and over multi-slot horizons: single-modality radar~\cite{demirhan2022radar,
luo2023radar}, vision~\cite{charan2022vision, jiang2022cv} and
LiDAR~\cite{klautau2019lidar, jiang2023lidar}, multimodal
fusion~\cite{Tian2023MultimodalTF, shi2024multimodal, zhu2025crossmodal}, public
benchmarks on real measurements~\cite{charan2022challenge}, and the framing of
integrated sensing and communication (ISAC)~\cite{cheng2022isac, yi2024beam}. ISAC predictive beamforming tracks a vehicle from radar echoes and outperforms feedback-based beam tracking in simulation~\cite{liu2020radar, yuan2021bayesian}.
Efficiency work compresses these predictors, frequently by knowledge distillation~\cite{ma2025kd, park2025ckd, shang2025multimodal}. Robustness work extends them
to adverse conditions such as low light~\cite{wang2025lowlight} or to
environment-semantic representations~\cite{yang2023env, imran2024env}.

Across the prediction studies summarized in Table~\ref{tab:positioning}, the sensor is
typically invoked \emph{unconditionally}, on every decision whether or not that decision
needs it. The baseline is exhaustive or hierarchical \emph{beam training} rather than the
link's own power feedback. Because those entries carry no measured feedback-only baseline, they do not report the incremental value of sensing over what
the link already knows. The key novelty of this work is therefore not the sensing model, which we take from our companion study (Sec.~\ref{sec:arm}), but the feedback-referenced evaluation framework and the
incremental value of sensing it measures. We also present a controller that admits sensing
through probe verification and backs off after a failed trial. 

\begin{table*}[t]\centering
\caption{Positioning: who decides when to sense, from what signal, and against what
baseline. \ding{51}/\ding{55} = present/absent; ``n/a'' = no sensing arm exists to
price. Among the works summarized here, only ISAC predictive beamforming compares sensing with feedback-based tracking, in simulation; none reports the incremental value of switching sensing on relative to an explicit feedback-only tracker on measured data under explicit probing accounting.}
\label{tab:positioning}\scriptsize
\setlength{\tabcolsep}{3pt}
\begin{tabular}{@{}p{0.33\textwidth}p{0.21\textwidth}ccc@{}}
\toprule
\textbf{Work} & \textbf{When-to-sense signal} & \textbf{Sensing} & \makecell{\textbf{Feedback-only}\\\textbf{baseline}} & \makecell{\textbf{Measured value}\\\textbf{of sensing}}\\
\midrule
Sensing-aided prediction~\cite{demirhan2022radar, ma2025kd, wang2025lowlight} & none (always-on) & always-on & \ding{55} & \ding{55}\\
ISAC predictive beamforming~\cite{liu2020radar, yuan2021bayesian} & none (always-on echoes) & always-on & \ding{51} (simulated) & \ding{55}\\
Budget-scheduled sensing~\cite{zakeri2025drl, zakeri2025constrained, zakeri2025aoi} & budget queue $+$ data age & on-demand (only source) & \ding{55} & \makecell{rate--accuracy\\only}\\
\makecell[tl]{Uncertainty-aware probing $+$ shield\\\cite{probing2026adaptive}} & \makecell[l]{none; prior entropy\\sets probe count} & always-on & \ding{55} & \ding{55}\\
Camera-primed beam mgmt.~\cite{vibe2026} & SNR gates RF refinement & always-on (stage 1) & \ding{55} & \ding{55}\\
CommRad~\cite{commrad2024} & fixed recalibration period & always-on & \ding{55} & \ding{55}\\
AT-IEKF~\cite{atiekf2024} & channel-strength variability & n/a & n/a & n/a\\
\makecell[tl]{Uncertainty-triggered training\\\cite{hussain2021dual, thz2022event}} & \makecell[tl]{belief uncertainty /\\analytic error prior} & n/a & n/a & n/a\\
Geometric / learned timers~\cite{scirep2024beam, v2i2024rl} & geometry / kinematics & position or none & \ding{55} & \ding{55}\\
\midrule
\textbf{This work} & \makecell[tl]{\textbf{probe-verified trial;}\\\textbf{innovation for hand-back}} & model on demand & \ding{51} & \makecell{system gain,\\marginal value (dB)}\\
\bottomrule
\end{tabular}
\end{table*}

\subsection{Sensor Scheduling for Beam Management}
\emph{When} to acquire sensor data has been asked before, most directly by work that
schedules sensing under an average budget. A deep-Q or Lyapunov policy picks the
slots in which to acquire fresh data, and non-sensing slots reuse the stale
acquisition~\cite{zakeri2025drl, zakeri2025constrained, zakeri2025aoi}. Two
structural differences separate that formulation from ours. First, no serving-link
measurement decides whether the acquired data are used. The scheduling state holds the
current or cached sensing input, the virtual budget queue and the data age, but no
serving-link residual, probe outcome or tracker state, so beam selection relies on fresh
or stale sensing alone. Second, those works report
sensing-rate--accuracy tradeoffs rather than incremental value against a feedback-only tracker.

The adjacent mechanisms cited here likewise do not test an external sensor against the serving link.
They include geometric update timers driven by
position and speed~\cite{scirep2024beam}, learned alignment-interval policies keyed to
vehicle kinematics~\cite{v2i2024rl}, event-triggered acquisition in THz links driven by
an analytic error-growth prior~\cite{thz2022event}, and, in networked control,
value-of-information scheduling driven by a model-based control state rather than serving-link innovation~\cite{ayan2019aoi}. Threshold rules also have precedent in linear-quadratic-Gaussian ISAC control~\cite{isacfc2026}. Closest in spirit, a dual-timescale belief-state policy conditions
beam-\emph{training} effort on belief uncertainty~\cite{hussain2021dual}, though with
no external sensing arm and on synthetic channels. We differ on both axes at once:
admission is decided by same-epoch probes against a live feedback tracker, and the
value of sensing is then measured against the same controller with its sensor disabled.

\subsection{Feedback Tracking and RF-Checked Sensing}
\looseness=-1
Kalman-family beam trackers driven by pilot feedback are well established~\cite{va2016beam}, and
adapting the tracking cadence to link-quality variation has direct precedent. An
adaptive-timing iterated extended Kalman filter (AT-IEKF) stretches or shrinks its
update interval from the historical variation of post-beamforming channel
strength~\cite{atiekf2024}. As stated in Sec.~\ref{sec:feedback}, our tracker, a simpler
innovation-driven member of this mechanism class, is not a contribution of this paper; its role is to provide the feedback-only reference against which system gain is measured.
A study by the 3rd Generation Partnership Project (3GPP) evaluates machine-learning beam
prediction from partial or past beam measurements to reduce reference-signal overhead, without an
external sensor~\cite{3gpp38843}.

Hybrid systems combining sensing with RF feedback also exist. CommRad~\cite{commrad2024} recalibrates an always-running radar arm on a fixed schedule, while the camera-primed system of~\cite{vibe2026} measures a camera-proposed beam against an SNR target before accepting it and refines locally below that target. Uncertainty-aware probing adapts its probe count from
sensing-prior entropy, observes only selected beams, and applies an
absolute-SNR-plus-margin shield before locking~\cite{probing2026adaptive}. RF checking
before lock and adaptive probing are therefore not our claims. In those systems the
sensing arm still runs on every sweep, and where a check exists, acceptance is against a
fixed absolute threshold. Ours instead requires same-epoch measurements to show that the
candidate neighborhood outperforms the beam a live feedback tracker is serving. It also
suspends the arm after a failed check and carries each probe outcome into future tracker
state.

\section{System Model and Causal Evaluation}\label{sec:model}
\subsection{Measured Link and Beam Model}\label{sec:model_channel}
The DeepSense vehicle-to-infrastructure testbeds place a quasi-omnidirectional $60$-GHz transmitter on
the vehicle and a $16$-element uniform-linear-array receiver at the stationary base
station~\cite{10144504}. At decision epoch $t$, the receiver records power $P_t(b)$
for every $b$ in a fixed $C=64$ receive-combining codebook through an exhaustive
sweep. Our replay evaluates selection of these measured receive beams and calls the
selected action the \emph{serving beam}. The recordings are uplink; interpreting the
same indices as base-station downlink transmit beams requires calibrated reciprocity
and matched transmit/receive codebooks, neither of which the dataset validates.
Because every reward is read directly from $P_t(b)$, the evaluation assumes neither a single line-of-sight path nor an analytic or monotone beam pattern.

\subsection{The Decision Loop}\label{sec:model_loop}
Beam tracking is a sequential decision problem, not a per-frame classification
problem. Here a frame means one dataset \emph{decision epoch}, the native sampling interval of $111$\,ms in Scenario~9 and about $92$\,ms in Scenarios~31--33, not a 5G New Radio (NR) radio frame. Let $\mathcal{B}=\{0,\ldots,C-1\}$ denote the beam-index set. At $t=0$
link setup performs one full sweep and hands the policy its initial beam. At each
subsequent epoch $t$ the policy may perform three operations, listed without implying
an intra-epoch order:
(i) \emph{probe} a set $\mathcal{S}_t \subseteq \mathcal{B}$ of candidate beams,
observing their measured powers at cost $|\mathcal{S}_t|$;
(ii) \emph{invoke} its sensing model, obtaining a beam posterior from the sensor
window at a separately accounted invocation cost; and
(iii) \emph{commit} a serving beam $a_t$. The sensor is a co-located radar or camera
feeding a trained predictor (Sec.~\ref{sec:arm}). Replay permits at most one uncharged scalar
link-power read per epoch, on the incumbent or serving beam used by that epoch-level
decision. Every candidate evaluation in $\mathcal{S}_t$ is charged, including the incumbent when
$\mathcal{S}_t$ contains it. A probed candidate that is then committed for service is not
charged a second time; its measured power is reused.
The policy's information set is exactly its own past probe outcomes, these scalar link
observations, and sensor outputs, never the full power vector.

These operations specify information, cost and epoch-level state transitions, not NR
symbol timing. One probe is one additional candidate-beam power evaluation, the usual
unit of beam-training overhead, though not a one-to-one count of NR airtime or report
bits. In NR terms it is one candidate beam evaluated on an
uplink sounding-reference-signal occasion or, with beam correspondence, one
channel-state-information reference-signal resource in downlink P-2 beam
refinement~\cite{giordani2019tutorial}; the supplementary material details both
realizations. The design problem is to minimize the expected power loss under separate
probe and invocation budgets (Sec.~\ref{sec:model_voi}). Rather than solve it, we measure
where the two budgets are worth spending and give an adaptive policy.

\subsection{Objective and the Value of Information}\label{sec:model_voi}
Performance is the capture ratio $\rho_t = P_t(a_t)/\max_b P_t(b)$, reported as
mean power loss against the recorded-sweep oracle, alongside the two cost components
(additional candidate probes/epoch; model invocations/epoch). The average is taken
in decibels, $\mathbb{E}[-10\log_{10}\rho_t]$ over the scored frames, which differs from
$-10\log_{10}\mathbb{E}[\rho_t]$. Because $-10\log_{10}(\cdot)$ is convex, the figures we
report are the larger of the two and give weight to the occasional deep miss that outage
tracks. We report two quantities and keep them distinct:
\begin{align}
  G_{\mathrm{sys}} &= \mathcal{L}_{\mathrm{ref}} - \mathcal{L}_{\mathrm{ctrl}},\label{eq:gsys}\\
  V_{\mathrm{sense}} &= \mathcal{L}_{\mathrm{ctrl}}^{-\mathrm{s}} - \mathcal{L}_{\mathrm{ctrl}},\label{eq:voi}
\end{align}
where $\mathcal{L}_{\mathrm{ctrl}}$ is the controller's mean power loss. The reference loss
$\mathcal{L}_{\mathrm{ref}}$ is that of an explicitly specified feedback-only reference: the plain tracker of
Sec.~\ref{sec:feedback}, or a validation-selected re-sweeping or lock-loss-recovery tracker. Finally,
$\mathcal{L}_{\mathrm{ctrl}}^{-\mathrm{s}}$ is the loss of the same controller with its sensor disabled and every other
component, including the post-setup re-sweep, held fixed.
$G_{\mathrm{sys}}$ is the controller's \emph{system gain}. $V_{\mathrm{sense}}$, the
\emph{marginal value of sensing}, is the sensor's value of information to this controller
in a given operating regime and can be positive, zero or negative. The probe saving
$\Delta P_{\mathrm{sense}}$ is the sensor-off minus the sensor-on probes/epoch in the comparison that defines $V_{\mathrm{sense}}$. Section~\ref{sec:feedback} builds the plain reference.

A positive system gain is credited to the controller only
when it also spends no more on candidate-beam probes than the reference. Probing harder is itself a way to buy link quality, so a comparison ignoring cost
would credit gains that extra probes could have bought. In Scenario~9, where the marginal
value of sensing is not resolved at the native interval, the sensor's contribution is the
probe saving $\Delta P_{\mathrm{sense}}$ (Sec.~\ref{sec:voi_day}).

We also report the probing-cost ratio at which a feedback-only policy \emph{matches}
the controller's link quality (Sec.~\ref{sec:voi_frontier}). Sensor invocations are
reported alongside rather than folded in, since invocations and probes are not
commensurable without platform assumptions.

\subsection{Replay Protocol}\label{sec:replay}\label{sec:replay_protocol}
DeepSense~6G~\cite{10144504} records, at every epoch of every vehicle pass, the
received power of all $64$ receive-codebook beams, measured by an exhaustive sweep
at the base station. We replay these recordings as the closed loop of
Sec.~\ref{sec:model_loop}. Link setup at $t=0$ is one full sweep whose cost we do not
count; thereafter the policy is \emph{action-restricted}, reading at most one uncharged
incumbent/serving-beam power and only the candidate powers it explicitly probes. Fig.~\ref{fig:scenario_replay}(b) separates these policy-visible
observations from the full $64$-beam vector, which serves solely to score $\rho_t$
and define the recorded-sweep oracle after initialization.

This is a sample-level counterfactual replay. It assumes an
exogenous propagation trajectory and treats the entries of each exhaustive sweep as
belonging to the same decision epoch. It therefore evaluates the information loop at
the dataset sampling granularity, not intra-sweep signaling or action-dependent network dynamics. In particular, a candidate can be probed, selected,
and applied within the same epoch; measurement, reporting, and
beam-application delays are not modeled. Probe verification benefits from this timing: an admitted beam is scored with the sweep
that admitted it. Section~\ref{sec:voi_tail} tests one-epoch sensor, probe-outcome, and commitment
delays, and stale link observations; the remaining timing limits are stated in the supplementary material. The noise-free replay is deterministic given the cached predictions and the policy configuration.

\subsection{The Oracle-Feedback Flaw}\label{sec:replay_flaw}
The learned predictors in our companion study take only sensor inputs~\cite{precursor}, but its windowed persistence and linear reference baselines
are initialized from recorded optimal-beam labels in every window. Treating such a
reference as a deployable tracker silently grants a reacquisition at every window
boundary. With that oracle initialization, persistence on Scenario~32 loses just $0.17$\,dB despite a horizon-averaged top-$1$ accuracy of $43.8\%$, and it looks locked to the oracle. Its causal rollout
loses $2.49$\,dB, $2.32$\,dB more.
Scenarios~33, 31 and~9 repeat the pattern ($0.25$ versus $3.07$, $0.30$ versus $1.63$ and
$4.11$ versus $6.35$\,dB). A deployed policy must recover this re-anchoring
information through probing. In our replay the policy's own past decisions determine
which powers it observes next, so tracking errors compound as they would
in deployment. Weak link-setup sweeps add a further, site-specific source of optimism,
quantified in Sec.~\ref{sec:voi_day}.

\subsection{Scenarios, Splits, and Statistical Units}\label{sec:replay_data}
Fig.~\ref{fig:scenario_replay}(a) shows representative camera views of the four
scenarios and their site relationships.
Scenario~9 (daytime; the high-innovation regime of
Sec.~\ref{sec:feedback_results}) is split
chronologically at pass boundaries into $91$ training, $17$ validation and $28$ test passes.
Every learned component is trained on training passes only, with checkpoints selected on
validation, and results are reported on the test passes. The test passes were not held
back from sensing design, however. Before a validation split existed, the companion study
developed its encoders and training recipe with these passes as the only held-out data,
and it chose the radar student's input pairing on a pilot read on them~\cite{precursor}.
Three further scenarios come from two other base-station sites on urban streets.
Scenario~31 (daytime) was recorded on the same avenue as Scenario~9, from a separate
base station about $110$\,m away, and follows the same split protocol ($35/6/11$ training/validation/test passes). Scenarios~32 and~33 are the
\emph{same} road recorded twice, by day and by night. Scenario~32 is split $10/2/3$ and all $18$ Scenario-33 passes are test,
replayed with the Scenario-32-trained sensing models for a same-site day-to-night comparison. The pair shares the site and testbed but not the session: traffic,
occlusions and vehicle trajectories differ between the two recordings, so it tests day-to-night generalization rather than illumination in isolation.
Scenarios~9, 31 and~32 are daytime recordings at different sites. We selected
Scenario~31 expecting a second high-innovation site, but the measurements show otherwise
(Sec.~\ref{sec:feedback_results}). This unexpected result, together with the same-site
pair, allows us to ask whether lighting or link predictability separates
the regimes. About $1\%$ of frames outside Scenario~9 lack one or more beam readings. Serving such a beam scores the replay's per-frame loss cap of $60$\,dB (capture ratio
floored at $10^{-6}$); excluding these frames shifts no gain by more than $0.08$\,dB (supplementary material).

Because frames within a pass are strongly dependent, the pass is the statistical
unit. A policy's loss is its mean over all scored frames, and confidence intervals are
\emph{paired} pass-level bootstrap intervals on the difference between two policies
(supplementary material). They cover pass-to-pass variation of the fixed final policies,
not of the design choices behind them.

\subsection{Metrics and the Decibel Scale}\label{sec:replay_metrics}
To the metrics of Sec.~\ref{sec:model_voi} we add, for tail behavior, the
\emph{capture-loss exceedance}: the fraction of frames whose loss against the
oracle exceeds a stated threshold. Being oracle-relative, it is not an
outage probability; the outage quoted from prior work~\cite{vibe2026} uses a
quantile-defined SNR threshold and is not directly comparable. Top-$1$ accuracy is not a headline metric
here. On the scored Scenario-9 test frames the peak is flat: the runner-up beam sits within $0.23$\,dB of the best at the median and is an immediate neighbor $87.0\%$ of the time, consistent with the companion study~\cite{precursor}. One-bin errors are therefore nearly free in power while distant
errors are catastrophic, structure that accuracy ignores and decibels capture.

The same study relates accuracy to loss on the Scenario~9 test windows: one
percentage point of seven-slot top-$1$ accuracy corresponds to about $0.06$\,dB of mean loss, a descriptive slope rather than a calibration constant. The flat-peak statistic motivates the $\pm1$ verification probes of Sec.~\ref{sec:voi}. Every headline loss, gain and value in this paper is instead computed directly from measured per-beam powers in causal replay; none is
obtained by applying the companion's fitted slope.

\section{The Feedback and Sensing Arms}\label{sec:arms}
\subsection{Innovation-Driven Tracker in Beam-Index Space}\label{sec:feedback}\label{sec:feedback_tracker}
We use the following feedback-only tracker as the plain reference. It maintains the last acquired beam $b_{\mathrm{acq}}$ and its power $p_{\mathrm{acq}}$, a
constant-velocity estimate $\hat{v}$ of beam-index drift, and an innovation scale $\hat{\sigma}$, the last two updated by exponential averaging from
its own probe outcomes. At
frame $n$ after the last acquisition the predicted beam is $c = b_{\mathrm{acq}} + \hat{v}\,n$, rounded to a codebook index, and the positional uncertainty is
$u = \max(\hat{\sigma}, \sigma_{\min})\,n$. Uncertainty therefore grows at the
\emph{measured} innovation rate rather than an assumed one, which lets one
constant set serve very different regimes. The tracker serves on $c$ without
additional probes while
$u \le \delta$. It probes a window of half-width $W=\operatorname{clip}(\lceil u\rceil e,1,W_{\max})$ about $c$ when
$u > \delta$, when its received signal strength drops more than $\eta$\,dB below
the last acquisition (an \emph{alarm}), or on a slow periodic refresh. An acquisition takes the strongest probed beam $b$ and, with smoothing $\lambda=0.3$ and innovation offset
$\sigma_{\mathrm f}=0.1$, updates
\begin{align}
\hat{\sigma} &\leftarrow (1-\lambda)\,\hat{\sigma} + \lambda\Bigl(\tfrac{|b-c|}{\max(n,1)} + \sigma_{\mathrm f}\Bigr),\label{eq:sigma}\\
\hat{v} &\leftarrow (1-\lambda)\,\hat{v} + \lambda\,\tfrac{b-b_{\mathrm{acq}}}{\max(n,1)},\label{eq:vel}
\end{align}
before setting $b_{\mathrm{acq}}\gets b$, $p_{\mathrm{acq}}\gets P_t(b)$ and $n\gets0$, starting from
$\hat{v}=0$ and $\hat{\sigma}=\sigma_0=1.5$. The offset $\sigma_{\mathrm f}$ keeps $\hat{\sigma}$ from collapsing to zero when predictions are exact.
A reacquisition fails when the best probed
power lands more than $2\eta$\,dB below the last acquisition. It then raises an alarm and doubles the escalation factor $e$, which a successful acquisition resets to $1$, rather than searching the same stale location again.

Its remaining constants (Algorithm~\ref{alg:controller}) are shared across scenarios.
The one free parameter is the slack $\delta$, the positional uncertainty tolerated before
probing. It is selected per scenario as the lowest-loss value on that scenario's validation
passes: $0.4$ on Scenario~9, $0.6$ on Scenario~31 and $0.8$ on Scenario~32. Scenario~33,
the same road at night without a validation split, inherits $0.8$ (supplementary
material). The Scenario-9 reference is thus
fixed before its test passes are read; it loses $1.045$\,dB at $3.64$ probes/epoch.

We do not claim this mechanism as novel; it differs from its precedent (Sec.~\ref{sec:related}) only in state space, control signal and evaluation substrate. Because the controller of
Sec.~\ref{sec:voi} consumes only its serving beam, probe outcomes and innovation statistic, any tracker that produces them, including a learned one, can be substituted.

\subsection{Baseline Performance Across Regimes}\label{sec:feedback_results}
We first place the tracker among feedback-only baselines on the Scenario~9 test
passes, with every free parameter chosen on the validation passes and evaluated once
here. Fig.~\ref{fig:frontier} plots the family on the cost--loss plane. Causal
persistence keeps the link-setup beam and never probes; naive triggering probes around
the held beam only after the served power drops. Both fail ($6.35$ and $6.27$\,dB; not plotted): a
trigger that searches around a stale beam recovers almost nothing for its $0.55$ probes/epoch.
Fixed-cadence probing, which searches the same window about the current beam every
epoch, trades probing for quality directly, from
$5.51$\,dB at $\pm1$ to $0.80$\,dB at $\pm3$ and $6.76$ probes/epoch. The tracker
reaches $1.045$\,dB at $3.64$ probes/epoch, capturing at
least $90\%$ of the available power on $69.5\%$ of frames against $79.0\%$ for the $\pm3$ cadence at $1.9$ times the probing cost.

Because gains are measured against the reference, we also ask whether a
stronger \emph{estimator} would lose less. Two posterior-carrying families, a Bayesian filter and a banded hidden-Markov tracker,
were tuned on validation at each probing budget (supplementary material). Both are strong when they may probe freely: the Bayesian filter reaches $0.16$\,dB at
$10.4$ probes/epoch, substantially improving on the plain tracker. Both degrade sharply when probing is limited. At a budget of $3.6$ probes/epoch the better of the two spends $3.00$
and loses $5.85$\,dB, because a diffuse posterior needs measurements to sharpen. Below about $3$ probes/epoch the innovation tracker is therefore
the strongest feedback-only policy we found. Above that, a tracker with one scheduled
re-sweep after link setup is stronger (Sec.~\ref{sec:voi_day}), and the controller of
Sec.~\ref{sec:voi} adopts that re-sweep. The innovation trackers, with or without a re-sweep, are also the only policies that \emph{produce} an innovation estimate.

The other three scenarios differ sharply (Table~\ref{tab:voimatrix}). There the tracker
comes within $0.06$--$0.25$\,dB of the oracle while
probing $1.68$--$2.27$ candidate beams/epoch. With about two probes per epoch, the link's
own feedback already captures nearly all the available power. The composition of this group differs from what illumination alone would predict.
Only Scenario~33 is a night collection. Scenario~32 is the \emph{same road by day}
and falls in the same regime (innovation medians $0.36$ and $0.37$; tracker losses
$0.055$ and $0.249$\,dB). Scenario~31 lies $110$\,m up the same avenue,
away from Scenario~9's intersection. It also falls in the low-innovation group (median $0.50$, tracker $0.109$\,dB, causal persistence $1.63$\,dB where
Scenario~9 loses $6.35$). Three daytime scenarios therefore split across both
regimes, while a day/night pair at one site stays within one. What separates them is
how much of the optimal beam's motion the link fails to anticipate, which the tracker
measures directly. For beam \emph{tracking}, the day/night contrast that motivates
much sensing-robustness work does not separate the regimes, at least in the one same-road pair, where traffic and trajectories also differ.

\begin{remark}[The low-innovation ceiling]\label{rem:ceiling}
On the three low-innovation scenarios (two daytime, one night), feedback-only tracking leaves a mean loss of $0.06$--$0.25$\,dB at the native interval. \emph{No} sensing modality, learned or perfect, can therefore reduce mean loss there by more than that.  The power side of the value-of-information question is thus
answered for the low-innovation regime before any sensing model is introduced. The
relevant question there is whether any sensor has room to help, not how much a
particular sensor helps.
\end{remark}

\subsection{Regime Separation in the Innovation Statistic}\label{sec:feedback_hinge}
The same replays expose why a hand-back rule is possible at all. The tracker's innovation
statistic separates the regimes cleanly at the median: $0.91$ on Scenario~9 against
$0.50$, $0.36$ and $0.37$ on the low-innovation three. The split follows the site geometry. In Scenario~9 the optimal beam's net index change is $12.4$ indices/s,
about $60$ of the $64$ indices per pass (medians over passes). The dataset's calibrated
vehicle positions put the vehicle a median $11$\,m from the base station ($16.4^\circ$/s). At the other sites the net change is
$1.6$--$3.8$ indices/s, and uncalibrated GPS, indicative only, puts the vehicles $30$--$42$\,m away
($2.0$--$2.7^\circ$/s). The validation passes show the same split (supplementary material). Computed as a
byproduct of tracking, the innovation is the quantity Sec.~\ref{sec:voi} thresholds to
decide when to hand the link back to feedback and whether to retry the sensor. The medians separate but the upper tails do not, so every low-innovation scenario still produces innovations above the trial threshold. A policy
acting on this signal alone would therefore invoke its sensor where the sensor performs poorly (Sec.~\ref{sec:voi_need}).

\subsection{Invocation Cost and What an Invocation Delivers}\label{sec:arm}\label{sec:arm_arch}
\label{sec:arm_velocity}\label{sec:arm_kd}%
The sensing arm itself (architecture, training protocol, input ablations and
distillation study) comes from our companion study~\cite{precursor}.
Here we convert that arm's beam choices into measured received power on the
Scenario-9 test windows; the feedback baselines, replay protocol, controller and value
measurements appear only in this paper.

The sensing arm is a learned beam predictor. An encoder consumes a short window of
sensor frames, a recurrent temporal model with positional attention summarizes it,
and a head emits a beam posterior over the codebook; the controller consumes only
the current-slot output. Unless noted, policy results use the \emph{vision} arm, which on
Scenario~9 reaches $44.7\%$ current-slot top-$1$ and loses $0.396$\,dB on the frames it covers; the radar alternative reaches $37.4\%$ and $0.721$\,dB. The covered frames, or
\emph{anchors}, are those with a complete eight-frame sensor window. Over the full seven-slot horizon the companion study
measures the same architectures at $0.44{\pm}0.08$\,dB for vision and $0.64{\pm}0.08$
for the radar student~\cite{precursor}; we take both arms as given and price what
invoking them is worth. These losses should be read against the $0.06$--$0.25$\,dB ceiling of
Remark~\ref{rem:ceiling}. At the low-innovation sites the arm alone loses $0.47$--$2.49$\,dB on the frames it covers (supplementary material), more than sensing could win there.

Vision has ${\sim}1.8$\,M parameters and needs ${\sim}137$ million multiply--accumulates
per call, against the radar student's ${\sim}90$\,K and ${\sim}156$ million~\cite{precursor}.
Single-sample model inference takes
$4.4$ and $2.4$\,ms on a laptop-class CPU ($2.5$ and $2.1$ on its GPU). These
model-side times are small relative to the $92$--$111$-ms decision epoch, but they exclude
capture, preprocessing, radio measurement, reporting, and beam application. The
controller avoids the inference cost outside trials and accepted episodes: the model is idle on $93$--$99\%$ of
epochs in the low-innovation scenarios and on $67\%$ of them in the
high-innovation one. 

\emph{What an invocation is.} An invocation is one run of the sensing model, not a capture
event. The model needs a short window of past frames whenever it runs, so \emph{capture} is
treated as continuous, with the sensor filling a short ring buffer; its energy and latency
are not measured here. \emph{Inference} is where the parameters, multiply--accumulates and
latency above are spent, and the invocation fractions we report count exactly the frames on
which the model runs.

\subsection{Degradation at the Other Sites}\label{sec:arm_night}
These figures describe Scenario~9 only. Retrained on the other sites, the same architectures are far weaker. The deployed vision seed's current-slot top-$1$ on the replayed anchors is $27.9\%$, $7.0\%$ and $2.8\%$ on Scenarios~31, 32 and~33. The radar student collapses to a constant-beam predictor on Scenarios~32 and~33 (supplementary material). Scenarios~31 and~32 were
recorded in daylight, so illumination alone does not explain the degradation; scene
content and the smaller training sets ($35$ and $10$ passes against $91$) remain
confounded. The same-site pair adds one day-to-night observation. Replayed on that road at night,
the same Scenario-32-trained vision arm loses a further $1.0$\,dB on the frames it covers
($1.49 \to 2.49$\,dB). This is consistent with reports that vision-aided prediction degrades in low light~\cite{wang2025lowlight},
although traffic and trajectories also differ between the two sessions. Both
degradations occur where, by Remark~\ref{rem:ceiling}, sensing has little loss to recover.

\subsection{Model-Side Confidence}\label{sec:arm_signals}
The controller of Sec.~\ref{sec:voi} admits sensing on a \emph{link-side} test. The
obvious alternative is the sensing model's own confidence. For the
deployed vision arm, however, the softmax maximum and the top-two margin separate its large
errors with an area under the receiver operating characteristic curve (AUC) of only
$0.45$--$0.65$ on the replayed anchors (supplementary material). Section~\ref{sec:voi_attrib} tests
confidence directly as an admission rule. The tracker's innovation instead measures
disagreement against the \emph{measured link} and separates the regimes at the median
(Sec.~\ref{sec:feedback_hinge}). The controller therefore uses it as its hand-back signal
and leaves admission to the probe test.

\section{Proposed Controller: Probe-Verified Sensing Invocation with Blacklist Backoff}%
\label{sec:voi}\label{sec:voi_controller}
The controller treats each sensor proposal as a claim that the link can check cheaply.
In a short trial of at most two epochs, it probes the proposed beam and its two neighbors,
three measurements per epoch and a small fraction of a $64$-beam sweep. It admits the
sensor only if the best of them beats the served beam by a fixed margin. Once admitted,
the sensor's later predictions are served and rechecked periodically or after an alarm.
A failed trial suspends the sensor for a long blacklist period, so a sensor that is weak
at a site costs few probes and model calls there. Whenever sensing is not admitted, feedback tracking keeps
the link, and one coarse re-sweep shortly after link setup protects it against a weak
setup sweep. The principle is that \emph{trust is never free}: sensing receives the link only
after measured link power shows its proposal beats what feedback delivers, and loses it
when a recheck detects a power collapse.

\begin{figure}[t]
\centering
\includegraphics{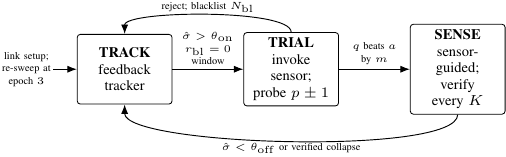}
\caption{Controller state transitions. A coarse re-sweep three epochs after link setup
restarts the tracker (left arrow); with the sensor disabled, the controller stays in
\textsc{track} with this re-sweep, the reference for the marginal value of sensing. A trial opens when the innovation estimate
exceeds $\theta_{\rm on}$ with no blacklist active. The restarted estimate usually meets this condition at the first covered frame; later
trials need it again. The sensor receives the link only after its probed neighborhood beats the incumbent by
the margin $m$ within a trial of at most two epochs; a failed trial imposes an
$N_{\rm bl}$-epoch blacklist. Accepted sensing is reverified every $K$ epochs; a
verified collapse hands the link back with a blacklist, and innovation below $\theta_{\rm off}$ or a lost sensor
window hands it back without one.}
\label{fig:controller_flow}
\end{figure}

\looseness=-1 The controller is the three-state feedback loop in Fig.~\ref{fig:controller_flow},
built on the tracker of Sec.~\ref{sec:feedback} with fixed slack $\delta=0.6$. Three
epochs after link setup it runs one charged coarse re-sweep (every fourth beam, then
$\pm2$ about the strongest, at most $20$ probes). It restarts the tracker at the best swept beam,
as beam refinement after initial access would. This removes the lock loss that a weak
setup sweep causes, in which the tracker stays about a dozen indices off the best beam
(Sec.~\ref{sec:voi_day}). In \textsc{track}, an available sensor window, no active blacklist, and innovation above $\theta_{\mathrm{on}}=0.8$ open a
sensing \emph{trial}, not an immediate switch. Because the re-sweep restarts the
innovation estimate at $\sigma_0=1.5$, the condition usually holds at the first covered
frame, so the first trial opens there. Later trials need the same conditions
(Sec.~\ref{sec:voi_attrib}). A trial lasts at most $L=2$ epochs. In each, the controller invokes
the model, keeps serving on its extrapolated beam, and probes the predicted beam $p$ and
its two immediate neighbors in place of the tracker's own probes. This $\pm1$ width follows the data: the runner-up beam neighbors the best beam $87.0\%$ of the time (Sec.~\ref{sec:replay_metrics}). The sensor enters \textsc{sense} only if the best probed candidate
$q\in\{p-1,p,p+1\}$ beats the served beam $a$ by the margin $m=0.5$\,dB,
\begin{equation}
P_t(q) \ge P_t(a)\,10^{m/10}.\label{eq:admit}
\end{equation}
If neither trial epoch admits it, the incumbent remains and the sensor is blacklisted
for $N_{\rm bl}=256$ epochs (counter $r_{\rm bl}$), longer than any validation or test pass of Scenarios~9 and~31. In
\textsc{sense} the model's prediction is served and reverified by $\pm1$ probing every
$K=4$ epochs or after an alarm. A verified collapse, $P_t(q)<p_{\mathrm{acq}}\,10^{-2\eta/10}$,
returns control to feedback with a blacklist; $\hat{\sigma}<\theta_{\mathrm{off}}=0.5$ or a lost sensor window (in either state) returns it without one. Every committed acquisition
updates the tracker's innovation and velocity states. In \textsc{sense} rechecks, \eqref{eq:sigma} uses
$p$ in place of $c$, so $\hat\sigma$ tracks the sensor's deviation. Algorithm~\ref{alg:controller}
gives the complete rule and constants.

One constant set (Algorithm~\ref{alg:controller}) serves all four scenarios. It was
selected on the \emph{validation} passes of Scenarios~9, 31 and~32. The search grid gives
the slack and the re-sweep's type and timing the same values as the feedback references.
It also spans the controller's own constants ($\theta_{\rm on}$, $\theta_{\rm off}$, $m$,
$L$, $N_{\rm bl}$, $K$). The selection rule,
recorded before the search, minimizes the mean \emph{regret}: the controller's
validation loss minus that of each scenario's own validation-selected feedback-only
tracker. Only settings that probe no more than that tracker on validation, in every
scenario, are eligible. The selected set was evaluated on test only after this selection, and every result in
Sec.~\ref{sec:results} uses it. The re-sweep itself entered the design after a test
analysis of an earlier constant set, after which all constants were re-selected on
validation (Sec.~\ref{sec:discussion}). Constants selected with Scenario~9 held out still
gain $+0.73$\,dB ($[0.23,1.28]$) over the plain tracker there. Holding out Scenario~31
or~32 instead gives $-0.10$\,dB ($[-0.30,0.07]$) and $-0.01$\,dB on the held-out site, and
two further training seeds of the vision arm both give $+0.79$\,dB on Scenario~9. The
supplementary material gives the grid and the selection chronology.
The controller is deliberately a threshold rule rather than a learned policy.

\begin{algorithm}[!htbp]
\small
\caption{Probe-verified sensing invocation with blacklist backoff and a post-setup
re-sweep. Boldface marks model invocations.}
\label{alg:controller}
\begin{algorithmic}[1]
\REQUIRE $\mathcal B{=}\{0,\ldots,C-1\}$;
$\mathcal N_{\mathcal B}(p){=}\{p-1,p,p+1\}\cap\mathcal B$;
  $\theta_{\rm on/off}{=}0.8/0.5$, $m{=}0.5$\,dB, trial length $L{=}2$,
  $N_{\rm bl}{=}256$ epochs, $K{=}4$, re-sweep epoch $k_{\rm rs}{=}3$;
$\delta{=}0.6$, $\eta{=}2$\,dB, refresh $T{=}16$, $W_{\max}{=}10$,
$e_{\max}{=}8$, $\sigma_0{=}1.5$, $\sigma_{\min}{=}0.2$,
$\sigma_{\mathrm f}{=}0.1$, averaging weight $\lambda{=}0.3$.
\STATE Full-sweep setup: $b_{\rm acq}\gets\argmax_bP_0(b)$,
 $p_{\rm acq}\gets P_0(b_{\rm acq})$, $\hat v\gets0$,
 $\hat\sigma\gets\sigma_0$, $n\gets0$, $e\gets1$,
  $a\gets b_{\rm acq}$, $k_{\rm chk}\gets k_{\rm tr}\gets r_{\rm bl}\gets0$,
 $\mathit{fire}\gets\mathrm{false}$, mode $\gets\textsc{track}$.
\STATE \textsc{Acquire}$(q)$ commits measured $q$ and sets $a\gets q$, updates
$(b_{\rm acq},p_{\rm acq},\hat v,\hat\sigma)$, resets $n$, and sets
$e\gets\min(2e,e_{\max})$ and $\mathit{fire}\gets\mathrm{true}$ after a drop exceeding $2\eta$,
else $e\gets1$ and $\mathit{fire}\gets\mathrm{false}$.
\FOR{each decision epoch $t$}
  \STATE $r_{\rm bl}\gets\max(r_{\rm bl}-1,0)$.
  \IF{$t=k_{\rm rs}$}
    \STATE probe every fourth beam and $\pm2$ about the strongest; restart the setup
    state on the best probed beam; mode $\gets\textsc{track}$; next epoch.
  \ENDIF
  \STATE $n\gets n+1$;
  $c\gets\operatorname{clip}_{\mathcal B}(\operatorname{round}(b_{\rm acq}+\hat vn))$;
    $u\gets\max(\hat\sigma,\sigma_{\min})n$.
  \IF{mode $\neq\textsc{track}$ \AND no sensor window} \STATE mode $\gets\textsc{track}$ (no blacklist). \ENDIF
  \IF{mode $=\textsc{track}$}
    \IF{$\hat\sigma>\theta_{\rm on}$ \AND $r_{\rm bl}=0$ \AND sensor window available}
            \STATE mode $\gets\textsc{trial}$; $k_{\rm tr}\gets0$.
    \ELSE
      \STATE if $\mathit{fire}$ \OR $u>\delta$ \OR $n\ge T$, probe the window of half-width
      $\operatorname{clip}(\lceil u\rceil e,1,W_{\max})$ about $c$ and \textsc{Acquire} its best;
      otherwise commit $c$ and set $a\gets c$.
    \ENDIF
  \ENDIF
  \IF{mode $=\textsc{trial}$}
        \STATE $k_{\rm tr}\gets k_{\rm tr}+1$; \textbf{run the sensing model}
    $\Rightarrow\boldsymbol\pi_t$; $p\gets\argmax_b\pi_t(b)$; $a\gets c$; probe
    $\mathcal N_{\mathcal B}(p)$; let $q$ be its measured best.
    \IF{$P_t(q)\ge P_t(a)10^{m/10}$}
      \STATE \textsc{Acquire}$(q)$; mode $\gets\textsc{sense}$; $k_{\rm chk}\gets0$.
    \ELSIF{$k_{\rm tr}=L$}
      \STATE commit $c$; mode $\gets\textsc{track}$; $r_{\rm bl}\gets N_{\rm bl}$.
    \ELSE \STATE commit $c$.
    \ENDIF
  \ELSIF{mode $=\textsc{sense}$}
    \STATE $k_{\rm chk}\gets k_{\rm chk}+1$; \textbf{run the sensing model}
    $\Rightarrow\boldsymbol\pi_t$; $p\gets\argmax_b\pi_t(b)$.
    \IF{$k_{\rm chk}\ge K$ \OR $\mathit{fire}$}
      \STATE Probe $\mathcal N_{\mathcal B}(p)$; let $q$ be its best and
      $z\gets[P_t(q)<p_{\rm acq}10^{-2\eta/10}]$; \textsc{Acquire}$(q)$ with $p$ in place of $c$
      in \eqref{eq:sigma};
      $k_{\rm chk}\gets0$.
      \IF{$z$} \STATE mode $\gets\textsc{track}$; $r_{\rm bl}\gets N_{\rm bl}$. \ENDIF
    \ELSE \STATE commit $p$ and set $a\gets p$. \ENDIF
    \IF{$\hat\sigma<\theta_{\rm off}$} \STATE mode $\gets\textsc{track}$. \ENDIF
  \ENDIF
  \STATE alarm: $\mathit{fire}\gets\mathrm{true}$ if $P_t(a)<p_{\rm acq}10^{-\eta/10}$.
\ENDFOR
\end{algorithmic}
\end{algorithm}

\section{Experimental Results}\label{sec:results}
\subsection{System Gain and the Marginal Value of Sensing}\label{sec:voi_day}
\begin{table*}[!t]\centering
\caption{Mean power loss and resource costs on test passes (causal replay,
vision arm). The controller uses one fixed parameter set across scenarios.
Radar-arm results appear in the supplementary material.}
\label{tab:voimatrix}\small
\setlength{\tabcolsep}{3pt}
\begin{tabular}{@{}lcccccccc@{}}
\toprule
 & \multicolumn{2}{c}{\textbf{S9 (high innov.)}} & \multicolumn{2}{c}{\textbf{S31 (low innov.)}} & \multicolumn{2}{c}{\textbf{S32 (low innov.)}} & \multicolumn{2}{c}{\textbf{S33 (low innov.)}}\\
\cmidrule(lr){2-3}\cmidrule(lr){4-5}\cmidrule(lr){6-7}\cmidrule(lr){8-9}
\textbf{Policy} & Loss & Cost & Loss & Cost & Loss & Cost & Loss & Cost\\
\midrule
Causal persistence & $6.35$ & $0$ & $1.63$ & $0$ & $2.49$ & $0$ & $3.07$ & $0$\\
Feedback-only tracker & $1.045$ & $3.64$ & $0.109$ & $2.27$ & $0.055$ & $1.68$ & $0.249$ & $1.77$\\
\quad with lock-loss recovery & $0.824$ & $3.83$ & $0.108$ & $2.42$ & $0.055$ & $1.68$ & $0.248$ & $1.82$\\
\quad with one scheduled re-sweep & $0.138$ & $5.40$ & $0.120$ & $2.95$ & $0.054$ & $1.95$ & $0.270$ & $2.10$\\
\quad\quad within $4$ probes/epoch & $0.370$ & $3.78$ & n/a & n/a & n/a & n/a & n/a & n/a\\
\textbf{Verified invocation (ours)} & $\bm{0.225}$ & $3.25$; $33.0\%$ & $\bm{0.118}$ & $2.30$; $6.8\%$ & $\bm{0.071}$ & $2.04$; $1.0\%$ & $\bm{0.186}$ & $2.12$; $1.5\%$\\
\quad sensor disabled & $0.260$ & $4.12$ & $0.115$ & $2.41$ & $0.052$ & $2.09$ & $0.163$ & $2.09$\\
Unverified invocation & $0.360$ & $1.34$; $84.7\%$ & $0.452$ & $0.39$; $95.0\%$ & $1.458$ & $0.18$; $97.8\%$ & $2.364$ & $0.34$; $92.5\%$\\
Always-on & $0.479$ & $0.55$; $85.2\%$ & $0.453$ & $0.21$; $95.3\%$ & $1.459$ & $0.10$; $97.8\%$ & $2.421$ & $0.12$; $97.3\%$\\
\quad $+$verification & $0.295$ & $3.11$; $85.2\%$ & $0.369$ & $3.07$; $95.3\%$ & $1.149$ & $3.03$; $97.8\%$ & $1.959$ & $3.04$; $97.3\%$\\
Oracle & $0$ & n/a & $0$ & n/a & $0$ & n/a & $0$ & n/a\\
\midrule
\textbf{Gain vs.\ feedback-only tracker} & \multicolumn{2}{c}{$+0.82$ $[0.30, 1.40]$} & \multicolumn{2}{c}{$-0.01$ $[-0.03, 0.02]$} & \multicolumn{2}{c}{$-0.02$ $[-0.07, 0.01]^{\ddagger}$} & \multicolumn{2}{c}{$+0.06$ $[-0.05, 0.24]$}\\
\textbf{Gain vs.\ re-sweep reference} & \multicolumn{2}{c}{$+0.15$ $[0.03, 0.27]^{\S}$} & \multicolumn{2}{c}{$+0.00$ $[-0.05, 0.09]$} & \multicolumn{2}{c}{$-0.02$ $[-0.07, 0.00]^{\ddagger}$} & \multicolumn{2}{c}{$+0.08$ $[-0.05, 0.31]$}\\
\textbf{Marginal value of sensing} & \multicolumn{2}{c}{$+0.04$ $[-0.06, 0.15]$} & \multicolumn{2}{c}{$-0.00$ $[-0.02, 0.02]$} & \multicolumn{2}{c}{$-0.02$ $[-0.06, -0.01]^{\ddagger}$} & \multicolumn{2}{c}{$-0.02$ $[-0.06, 0.00]$}\\
\bottomrule
\end{tabular}
\par\vspace{2pt}
\begin{minipage}{\textwidth}
\footnotesize\raggedright
\textit{Metrics.} Loss is in dB. The two cost entries are additional candidate-beam
probes/epoch and model-invocation percentage (trials plus accepted sensing).
Sensor-free policies list probes only. Gain rows give the reference's loss minus the controller's loss; the marginal value of sensing is the loss of the controller with its sensor disabled minus the controller's. Bold marks our controller.
\par
\textit{Comparison.} Feedback-only slack settings are specified in Sec.~\ref{sec:feedback}.
The lock-loss-recovery row adds a peak-relative power-drop test and a charged re-sweep, selected
on validation within $4$ probes/epoch. The scheduled re-sweep rows add one charged sweep
(all beams, or every fourth beam refined by $\pm2$) a few epochs after link setup and
restart the tracker. Slack, sweep type and delay are selected on validation, in
Scenario~9 also within $4$ probes/epoch ($^{\S}$the gain row uses the one selected within $4$ probes/epoch; against the
unconstrained one it is $-0.09$ $[-0.14,-0.04]$). The sensor-disabled row runs the
controller's own tracker and re-sweep without the model. Unverified invocation is the
controller without the admission trial, periodic verification or blacklist
(Sec.~\ref{sec:voi_need}).
Always-on serves the sensor's prediction, unprobed, on every frame with a complete
sensor window and elsewhere falls back on the plain tracker without the re-sweep; $+$verification also
probes the prediction's $\pm1$ neighborhood every epoch.
\par
\textit{Uncertainty.} Brackets are paired pass-level percentile bootstrap 95\% confidence intervals (CIs; unadjusted for multiplicity; $10^5$ resamples of complete passes with replacement), except $^{\ddagger}$Scenario~32: its bracket is the range across
three test passes. Its Table~\ref{tab:decim} results use the same three passes.
\end{minipage}
\end{table*}

We first compare the complete controller with feedback-only references, then isolate sensing by
disabling it while the controller's tracker and re-sweep stay unchanged.

\smallskip\noindent\textbf{The gain appears where feedback leaves loss.} In high-innovation
Scenario~9 the controller cuts the feedback-only tracker's loss from $1.045$ to
$0.225$\,dB, a system gain of $+0.82$\,dB, with fewer probes
(Table~\ref{tab:voimatrix}). In the three low-innovation scenarios the same constants
lose at most $0.02$\,dB to feedback-only and invoke the model on $1.0$--$6.8\%$ of frames,
whereas always-on sensing loses more than feedback-only in each of them. There the system gain
and the marginal value of sensing are indistinguishable from zero or slightly negative.
The controller also probes more than the plain tracker there, up to $22\%$ more on
Scenario~32, mostly for the re-sweep and the tighter slack.

\begin{figure}[t]\centering
\includegraphics{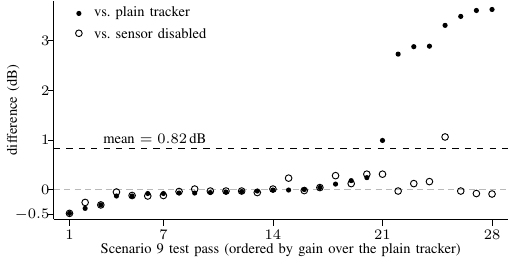}
\caption{Per-pass loss difference on the $28$ Scenario-9 test passes against the plain
tracker (filled) and against the controller with its sensor disabled (hollow), in the
same pass order. Positive values mean that the controller loses less; zero denotes parity; the dashed line is the frame-weighted mean gain over the plain tracker.}
\label{fig:perpass}
\end{figure}

\smallskip\noindent\textbf{Seven passes carry the gain.} Fig.~\ref{fig:perpass} shows
that seven of the $28$ test passes gain $2.73$--$3.63$\,dB over the plain tracker, while
the other $21$ average $-0.02$\,dB. The median pass therefore shows no gain: a sign test
does not reject parity ($p=0.57$), although a sign-flip permutation test on the mean
does ($p=0.009$).

\begin{figure}[t]\centering
\includegraphics{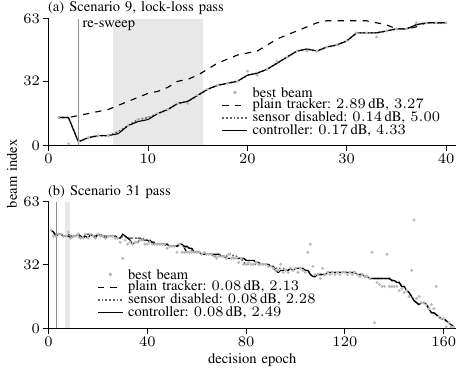}
\caption{Two test passes, each with the median marginal value of sensing in its group:
(a) among the seven Scenario-9 lock-loss passes (its system gain, $2.73$\,dB, is the smallest of the
seven), (b) among the eleven Scenario-31 passes. Legends give each policy's mean loss and
probes/epoch on the pass. Dots mark the best beam (isolated far dots are near-ties), shading marks
frames on which the model runs, the vertical line marks the re-sweep, and the solid line hides the
others where they coincide. (a) The plain tracker stays about a dozen indices off the best beam for
$30$ epochs. The re-sweep returns both controller variants to it; the admitted sensor then saves
$27$ probes at a $0.03$\,dB higher loss until a verified collapse at epoch~$15$ suspends it.
(b) The only trial fails because feedback already serves within the admission margin, and the
blacklist then holds; the trial also shifts the tracker, which probes more on this pass.}
\label{fig:example}
\end{figure}

\smallskip\noindent\textbf{Those passes lose lock at link setup.} Each begins with a
nearly flat setup sweep whose strongest beam lies $12$--$13$ indices from the beam that
dominates three epochs later. The plain tracker locks onto the setup beam and then follows the
moving best beam about a dozen indices off; its alarm, set relative to the weak setup
power, rarely fires. This happens on $22\%$ of all $136$
Scenario-9 passes; Fig.~\ref{fig:example}(a) shows one of them. One charged re-sweep a few epochs after setup, as beam refinement
after initial access would provide, removes every lock loss. A power-drop test relative to the last sweep's best power,
with a charged re-sweep (lock-loss recovery), fires too rarely for the same reason
(Table~\ref{tab:voimatrix}). Because the controller includes such a re-sweep
(Sec.~\ref{sec:voi_controller}), the re-sweeping trackers are the relevant references.
Against the one selected on validation within four probes/epoch, just above the plain
tracker's budget, the controller keeps $+0.15$\,dB with fewer probes. Of the two
validation-selected references, only the unconstrained one loses less, by $0.09$\,dB at
$66\%$ more probing.

\smallskip\noindent\textbf{Sensing and the re-sweep correct the same failure.} Switching the
sensor and the re-sweep independently, with every other constant fixed, separates their roles. Without either, the controller's tracker loses
$1.209$\,dB, all seven lock-loss passes included. Either component alone removes every
lock loss, but they spend different resources: the sensor alone reaches $0.419$\,dB at
$2.62$ probes/epoch, and the re-sweep alone $0.260$\,dB at $4.12$. Without the re-sweep,
the marginal value of sensing is thus $0.79$\,dB ($[0.33,1.29]$).

\smallskip\noindent\textbf{With the re-sweep in place, sensing saves probes rather than
reducing loss.} Adding the sensor to the re-sweep cuts probing from $4.12$ to $3.25$
probes/epoch, a saving of $21\%$ ($[13\%,29\%]$), while the loss moves from $0.260$ to
$0.225$\,dB. This marginal value of sensing ($+0.04$\,dB) is not resolved, with a one-sided
$95\%$ lower bound of $-0.05$\,dB. The two marginal values differ by $0.75$\,dB
($[0.34,1.22]$), consistent with both components correcting the same failure. With
both in place, sensing substitutes external information for candidate-beam measurements
at the native interval (Fig.~\ref{fig:example}(a)). Secs.~\ref{sec:voi_decim} and~\ref{sec:voi_tail} show where
it also reduces loss, and Sec.~\ref{sec:discussion} what the saving is worth. With $\pm1$
verification on every frame, always-on sensing, whose fallback tracker lacks the re-sweep, loses
$0.07$\,dB more than the controller at slightly lower probing, a difference that is not significant. It runs the model on
$85\%$ of frames, against $33\%$ (Table~\ref{tab:attrib}). With the radar arm the controller loses $0.301$\,dB on
Scenario~9 and stays near feedback-only at the low-innovation sites (supplementary
material), so the mechanism is not specific to vision.

\begin{figure}[t]\centering
\includegraphics{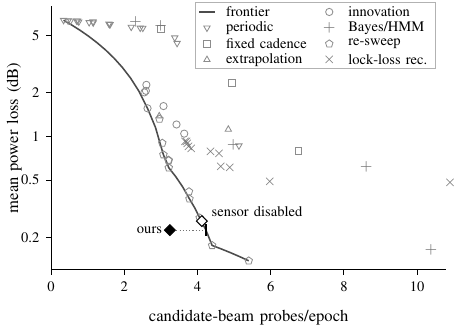}
\caption{Scenario~9 mean power loss versus candidate-beam probing cost; the loss
axis is logarithmic. Hollow markers and crosses denote feedback-only policies and the
solid curve their lower convex hull, attainable by time sharing between policies.
The filled diamond is our controller and the hollow diamond the same controller with
its sensor disabled, itself a re-sweeping feedback-only policy; the dotted segment ends
at the equal-loss frontier crossing, $4.23$ probes/epoch, $1.30$ times the controller's
budget ($95\%$ interval $[1.15,1.43]$ over pass resamples). Full-sweep policies lie beyond
the plotted range (supplementary material).}
\label{fig:frontier}
\end{figure}

\smallskip\noindent\textbf{Tested feedback-only policies need 1.3 times the
probes.}\label{sec:voi_frontier} Fig.~\ref{fig:frontier} compares the controller with
$205$ feedback-only policies: periodic probing of a $\pm W$ window every $N$ epochs,
fixed-cadence probing, trajectory extrapolation, innovation trackers, the same trackers
with lock-loss recovery or one scheduled re-sweep, and the posterior-carrying estimators
of Sec.~\ref{sec:feedback_results}. Because policies can be time-shared across passes,
the frontier is their lower convex hull. It reaches the controller's loss at
$4.23$ probes/epoch, $1.30$ times the controller's budget, or $1.20$ times after
amortizing each pass's setup sweep (supplementary material); at the controller's budget
it lies $0.37$\,dB ($[0.19,0.66]$) higher. The sensor-disabled controller sits just above the frontier.
Sensing thus does not reveal a beam that feedback could never find; at this site and
cadence it supplies the same information with fewer probes.

\subsection{Attribution of the Gain}\label{sec:voi_attrib}
The ablations separate three roles: current sensing information proposes a beam,
verification checks the proposal against measured link power, and the blacklist,
not the innovation gate, limits model calls and repeated trials. The candidate and gate
controls in Table~\ref{tab:attrib} change one component at a time; the verification
ablations below remove the admission trial alone or together with periodic verification
and the blacklist.

\smallskip\noindent\textbf{Non-informative proposals raise loss.} The candidate
controls keep the same neighborhood probing and acquisition rules and change only the
proposal (Table~\ref{tab:attrib}). In Scenario~9, replacing the sensor's proposal with the
tracker's own extrapolation raises loss by $1.93$\,dB, and with a random nearby proposal
by $2.92$\,dB. A prediction $20$ frames earlier, or the current one where none is that old ($58\%$ of covered
frames), raises loss by $0.17$\,dB, because verification rejects the stale proposal often enough. Verification
does not, however, stop a non-informative proposal from winning a trial once the served
beam is lost; \textsc{sense} then serves it until a scheduled check or an alarm.

Perfect proposals would add little at these settings. The oracle control replaces each
available sensor proposal with the frame's best beam, with the controller's settings
unchanged. It lowers the Scenario-9 loss by $0.05$\,dB, and $17\%$ of the feedback-only
loss stays unrecovered, because coverage, trial timing and verification still constrain
performance. In Scenario~32 the real sensor wins none of its trials, since the served
beam is already within the $m=0.5$\,dB margin of the optimum. Its marginal value of sensing
($-0.02$\,dB) thus comes from rejected trials, which displace the tracker's own probes.

\smallskip\noindent\textbf{The blacklist, not the gate, limits model
calls.}\label{sec:voi_calls} With verification unchanged and the blacklist removed, the
model runs on $30.1$--$65.9\%$ of frames across the four scenarios instead of
$1.0$--$33.0\%$. Loss on Scenario~33 rises from $0.186$ to $1.492$\,dB. An
eight-epoch blacklist already restores most of that loss. Removing the innovation gate
instead, so that a trial opens whenever coverage and the blacklist permit, changes loss
by at most $0.027$\,dB and invocations by at most $0.7$ percentage points. Random opening
at a matched rate gives the same Scenario-9 loss (Table~\ref{tab:attrib}).

\smallskip\noindent\textbf{The gate is nearly inert.} The re-sweep restarts the
innovation estimate above $\theta_{\mathrm{on}}$, so the first trial opens at the first
frame with a sensor window in most passes ($27$ of $28$ on Scenario~9). In Scenarios~9 and~31 a
failed trial or a verified collapse ends trials for the rest of the pass, because the blacklist outlasts the
pass (Fig.~\ref{fig:example}). In Scenario~9 the
controller runs $37$ trials in $28$ passes, one or two per pass, and wins $26$, so in
these short passes it acts as an early probe test of the sensor. Waiting for measured innovation
costs little once the re-sweep has removed the lock loss. A gate that reads a separate
innovation estimate started at zero loses $0.06$\,dB ($[-0.01,0.16]$), and one that waits
for six acquisitions loses $0.035$\,dB; neither difference is significant. Gates on measurement age cut invocations but
raise probing to the sensor-disabled level
(Table~\ref{tab:attrib}).

\smallskip\noindent\textbf{Verification alone does not make always-on sensing safe.}
Adding $\pm1$ verification on every frame or every second frame lowers the Scenario-9
loss of always-on sensing, and with verification on every frame it is not significantly
worse than the controller (Table~\ref{tab:attrib}). In the other three scenarios,
however, always-on sensing stays $0.26$--$2.17$\,dB behind feedback-only with or without
verification.

\smallskip\noindent\textbf{Confidence admission and sensor-prior fusion.} Two
further competitors use the model's posterior on every covered frame and are selected with
the controller's own rule (supplementary material). Confidence admission mostly declines
to admit the sensor and lowers the Scenario-9 loss by at most $0.07$\,dB relative to its
plain-tracker fallback. No sensor-prior fusion setting held to the reference's probing beats
feedback-only on validation. Both run the model on $85$--$98\%$ of frames. Neither has
the re-sweep, so their gaps to the controller do not isolate the admission rule.

\begin{table}[t]\centering
\caption{Attribution controls on Scenario~9 test passes. Candidate and gate controls
retain all of the controller's settings except the component identified in each row. The
first two rows reproduce the references in Table~\ref{tab:voimatrix}; the final block is
always-on sensing and has no trial. Positive loss differences mean that the controller
is better; brackets are paired pass-level bootstrap 95\% CIs.}
\label{tab:attrib}\scriptsize
\setlength{\tabcolsep}{2pt}
\begin{tabular}{@{}lccccc@{}}
\toprule
\textbf{Policy} & Loss & Probes & Model & p95 & $\Delta$ vs.\ ours\\
 & (dB) & /epoch & (\%) & (dB) & (dB, $95\%$ CI)\\
\midrule
Feedback-only tracker & $1.045$ & $3.64$ & $0$ & $5.53$ & $+0.82$ $[0.30, 1.40]$\\
\textbf{Verified invocation (ours)} & $\bm{0.225}$ & $3.25$ & $33.0$ & $\bm{1.14}$ & reference\\
\midrule
\multicolumn{6}{@{}l@{}}{\emph{candidate replaced (perfect sensing information)}}\\
\quad oracle candidate & $0.179$ & $2.86$ & $40.2$ & $0.97$ & $-0.05$ $[-0.08, -0.01]$\\
\midrule
\multicolumn{6}{@{}l@{}}{\emph{candidate replaced (no current sensing information)}}\\
\quad own extrapolation & $2.158$ & $3.66$ & $19.8$ & $9.08$ & $+1.93$ $[1.06, 2.87]$\\
\quad random offset & $3.145$ & $3.29$ & $31.5$ & $9.67$ & $+2.92$ $[2.04, 3.79]$\\
\quad stale prediction & $0.393$ & $3.34$ & $30.6$ & $1.90$ & $+0.17$ $[0.00, 0.41]$\\
\midrule
\multicolumn{6}{@{}l@{}}{\emph{gate replaced (everything else unchanged)}}\\
\quad random opening & $0.244$ & $3.28$ & $33.2$ & $1.13$ & $+0.02$ $[-0.03, 0.08]$\\
\quad measurement age ${\ge}2$ & $0.248$ & $3.98$ & $13.8$ & $1.30$ & $+0.02$ $[-0.05, 0.10]$\\
\quad measurement age ${\ge}3$ & $0.254$ & $4.11$ & $0.4$ & $1.34$ & $+0.03$ $[-0.06, 0.15]$\\
\quad no gate at all & $0.251$ & $3.30$ & $33.5$ & $1.13$ & $+0.03$ $[-0.03, 0.09]$\\
\midrule
\multicolumn{6}{@{}l@{}}{\emph{always-on sensing, with and without verification}}\\
\quad no verification & $0.479$ & $0.55$ & $85.2$ & $2.37$ & $+0.25$ $[0.14, 0.39]$\\
\quad $+$verify, period $2$ & $0.397$ & $1.81$ & $85.2$ & $2.03$ & $+0.17$ $[0.07, 0.30]$\\
\quad $+$verify, period $1$ & $0.295$ & $3.11$ & $85.2$ & $1.55$ & $+0.07$ $[-0.03, 0.19]$\\
\bottomrule
\end{tabular}
\\[2pt]{\footnotesize Model is the percentage of frames on which the sensing model
runs (verification trials plus sensing); p95 is the $95$th percentile of per-frame
capture loss; measurement age is the number of epochs since the tracker's last
acquisition ($n$ in Algorithm~\ref{alg:controller}), and an age gate opens a trial when it
reaches $2$ or $3$, in place of the innovation condition; period $1$ or $2$ means that the
always-on sensor's proposed neighborhood is probed on every frame or every second
frame.}\end{table}

\smallskip\noindent\textbf{Verification protects the low-innovation
sites.}\label{sec:voi_need} The first ablation removes the admission trial but keeps
periodic verification; the second also removes periodic verification and the blacklist,
so accepted sensing is re-probed only after an alarm (unverified invocation in
Table~\ref{tab:voimatrix}). The low-innovation sites need verification because their
innovation still crosses the threshold: the medians lie below
$\theta_{\mathrm{on}}=0.8$, but the 99th percentiles are comparable with Scenario~9's, so
the gate opens there too. Without the admission trial, their loss rises from
$0.07$--$0.19$\,dB to $0.30$--$1.35$\,dB, and without verification or the
blacklist to $0.45$--$2.36$\,dB. In Scenarios~31 and~33 every paired interval except the first ablation's on Scenario~31
excludes zero; every Scenario-32 pass is worse under both ablations. In Scenario~9 the two ablations cost $0.085$ and
$0.135$\,dB, and both intervals exclude zero.

The distinction is structural, not parametric: innovation in \textsc{track} measures how badly
feedback tracks, not whether the sensor is better. Verification supplies that
missing comparison. It enforces one necessary condition for admission, an immediate
measured improvement over the serving beam. It does not certify the sensor's quality or
timeliness over the decisions that follow; Secs.~\ref{sec:voi_decim}
and~\ref{sec:voi_tail} show where that matters.

\subsection{What Governs the Gain and the Value of Sensing}\label{sec:voi_decim}
\noindent\textbf{Decimation protocol.}
Site geometry, beam motion, and sensor accuracy vary together across scenarios.
For a within-scenario sensitivity test, each policy observes and decides only on every
$d$-th frame, $d\in\{1,2,4\}$, and holds its beam in between. A decision thus spans $d$ native
epochs ($d\times111$\,ms in Scenario~9), and every native frame is scored. The site, sensing model, codebook, and controller constants stay fixed, with
no retuning at any $d$. The feedback-only tracker likewise keeps its native slack;
re-selecting it at each $d$ does not lower its test loss where the controller gains
(supplementary material). In Table~\ref{tab:decim} the model still receives eight
consecutive native frames; this optimistic construction preserves its training sampling
interval. A pessimistic construction ($d$-spaced inputs) feeds the same checkpoint eight
frames spaced $d$ native epochs apart, without retraining, and so also decimates the sensor
history.

\smallskip\noindent\textbf{Longer intervals raise the gain most where the sensor is
accurate.} In Scenario~9 the gain over feedback-only rises from $+0.82$ to $+1.88$ and
$+3.66$\,dB at $d=2$ and $4$ (Table~\ref{tab:decim}). Feedback-only loss grows with the
beam step per decision, to $5.98$\,dB at $d=4$, against $2.32$\,dB for the controller. The
other sites, whose arms are weaker, gain little or lose, down to $-0.56$\,dB on Scenario~32
at $d=4$.
Verification does not prevent that harm: higher innovation opens more trials, and a weak
sensor can occasionally pass a margin check without providing sustained tracking accuracy.
The pessimistic construction cuts Scenario-9 current-slot accuracy from $44.7\%$
to $0.9\%$ at $d=4$ and reduces prediction coverage. Across the two input constructions,
the gain over the plain tracker ranges from $+0.73$ to $+1.88$\,dB at $d=2$ and from
$+0.89$ (at $13\%$ more probes, not credited; Sec.~\ref{sec:model_voi}) to $+3.66$\,dB at
$d=4$. These ranges are sensitivity cases, not confidence intervals.

\smallskip\noindent\textbf{A re-sweep does not replace full-rate sensor inputs.} Re-selected on
every-frame validation losses, the re-sweeping feedback reference loses $1.862$ and
$4.935$\,dB at $d=2$ and $4$ while probing $5.85$ and $11.55$ times per decision. With
consecutive sensor inputs the controller is $1.18$ and $2.62$\,dB better ($[0.63,1.82]$
and $[2.40,2.83]$) at $53\%$ and $37\%$ of the reference's probes per decision. With $d$-spaced inputs the two are not
significantly different (supplementary material). The best beam's median net displacement
is $1.4$, $2.9$ and $6.0$ indices per decision in Scenario~9 at $d=1$, $2$ and $4$, against
$0.16$--$0.38$ at the far sites at $d=1$. This motion is a proxy for the decision interval
relative to the beam coherence time~\cite{va2017impact}, and the gain over the strongest
reference rises with it, provided the sensor keeps its frame rate.

\begin{figure}[t]\centering
\includegraphics{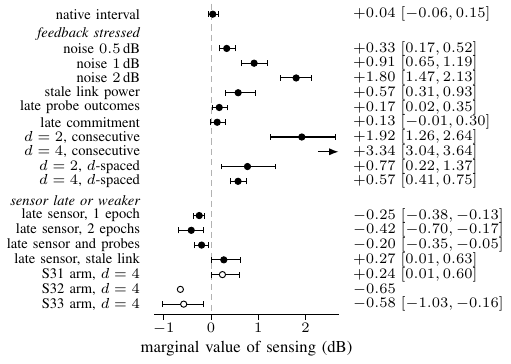}
\caption{Marginal value of sensing on test passes (loss of the controller
with its sensor disabled minus that of the controller; paired pass-level 95\% CIs; positive
means that sensing reduces loss). Filled: Scenario~9; hollow: other sites with their own
arms and consecutive inputs. Rows without $d$ are at the native interval. Noise: standard
deviation of an error on every observed power. Stale link power: every observed power is
one epoch old. Late probe outcomes: a probing epoch serves its pre-probe beam. Late
commitment: every committed beam is applied one epoch late. Late sensor: current-slot output one
or two epochs old; combined with late probes or stale link power, each is one epoch late.
$d$: decimation factor, with beams held between decisions and every frame scored;
consecutive inputs are eight consecutive native frames, $d$-spaced
inputs eight frames $d$ epochs apart (Sec.~\ref{sec:voi_decim}). The $d=4$
consecutive-input value is off scale; Scenario~32 (three passes) has no interval.}
\label{fig:vsense}
\end{figure}

\smallskip\noindent\textbf{When link feedback is stressed, an accurate sensor also
reduces loss.} Fig.~\ref{fig:vsense} collects the marginal value of sensing under the
stresses we applied.
On Scenario~9 it rises from $+0.04$\,dB at the native interval to $+1.92$ and $+3.34$\,dB
at $d=2$ and $4$ and stays positive with $d$-spaced inputs. Sensing also saves
$12$--$48\%$ of the probes at $d=2$ and $4$ (supplementary material). On Scenario~9, a longer decision
interval thus creates feedback loss that sensing partly recovers, even relative to the
same controller without sensing. This holds even with $d$-spaced inputs, whose
exact-index accuracy falls to $0.9\%$ at $d=4$; top-$1$ accuracy does not measure the
power value of a proposal that $\pm1$ probes then check (supplementary material). This is not universal: at $d=4$ the
weaker arm of Scenario~31 adds $+0.24$\,dB ($[0.01,0.60]$), and those of
Scenarios~32 and~33 cost about $0.6$\,dB, which probe verification does not prevent.

\smallskip\noindent\textbf{Interpretation and scope.}
Across the twelve scenario-by-decimation cells, the gain follows current-slot sensing
accuracy (Spearman $+0.72$) and innovation ($+0.53$) more closely than the drift of Table~\ref{tab:decim}
($+0.27$). The pattern is descriptive, since the cells reuse passes of four scenarios.
Decimation also changes feedback timing, prediction coverage and the sensor-to-commit
latency budget, so the experiment tests sensitivity to a longer decision interval, not
vehicle speed or beam dynamics alone (supplementary material).

\begin{table}[t]\centering
\caption{Decision-interval sensitivity with consecutive sensor inputs and fixed
controller settings. Each decision spans $d$ frame periods, over which its beam is held, and every frame is scored; drift is the mean absolute beam-index step per decision. Drift includes index jitter
around a flat power peak, so a low-innovation site can show more drift than Scenario~9
while its best beam barely moves (Sec.~\ref{sec:voi_decim}). Feedback-only is the plain tracker of Sec.~\ref{sec:feedback};
gain is its loss minus controller loss.}
\label{tab:decim}\small
\setlength{\tabcolsep}{3.5pt}
\begin{tabular}{@{}llccccc@{}}
\toprule
Scenario & $d$ & drift & innov. & feedback-only & controller & gain\\
 & & (bins) & median & (dB) & (dB) & (dB)\\
\midrule
\multirow{3}{*}{S9} & $1$ & $1.85$ & $0.91$ & $1.045$ & $0.225$ & $+0.82$\\
 & $2$ & $3.34$ & $1.44$ & $2.561$ & $0.681$ & $+1.88$\\
 & $4$ & $6.54$ & $1.60$ & $5.975$ & $2.317$ & $+3.66$\\
\midrule
\multirow{3}{*}{S31} & $1$ & $1.51$ & $0.50$ & $0.109$ & $0.118$ & $-0.01$\\
 & $2$ & $1.57$ & $0.58$ & $0.171$ & $0.169$ & $+0.00$\\
 & $4$ & $2.22$ & $1.08$ & $0.701$ & $0.389$ & $+0.31$\\
\midrule
\multirow{3}{*}{S32} & $1$ & $2.13$ & $0.36$ & $0.055$ & $0.071$ & $-0.02$\\
 & $2$ & $1.64$ & $0.41$ & $0.113$ & $0.085$ & $+0.03$\\
 & $4$ & $2.13$ & $0.64$ & $0.318$ & $0.876$ & $-0.56$\\
\midrule
\multirow{3}{*}{S33} & $1$ & $1.73$ & $0.37$ & $0.249$ & $0.186$ & $+0.06$\\
 & $2$ & $2.05$ & $0.53$ & $0.226$ & $0.278$ & $-0.05$\\
 & $4$ & $2.36$ & $0.73$ & $0.942$ & $0.961$ & $-0.02$\\
\bottomrule
\end{tabular}
\end{table}

\subsection{Robustness of the Operating Point}\label{sec:voi_tail}
\noindent\textbf{In Scenario~9 the controller removes most deep losses.} It cuts the
fraction of frames losing more than $3$\,dB from $16.7\%$ for the plain tracker to
$0.9\%$, mostly by correcting the lock-loss passes. In Scenario~33 always-on sensing
exceeds $3$\,dB on $38.2\%$ of frames, against $1.3\%$ for feedback-only and $1.6\%$ for
the controller (supplementary material). Scenario~32's three passes are too few for tail
estimates.

\smallskip\noindent\textbf{The operating point is robust to single-parameter
changes.}\label{sec:voi_robust} Of $68$ single-parameter perturbations, $17$ per scenario
with the re-sweep's delay and type included, $65$ stay within $0.1$\,dB of the selected
loss (supplementary material).

\smallskip\noindent\textbf{Noisy feedback makes sensing reduce loss.} On Scenario~9 we
add zero-mean Gaussian errors in dB, with standard deviation
$\sigma_{\mathrm{m}}\in\{0.5,1,2\}$\,dB, to every observed power ($20$ realizations each).
The errors are independent across frames and beams and shared by all policies; rewards
use the unperturbed powers, and no constant is retuned.
Feedback-only degrades faster than the controller, so the gain
over the plain tracker grows to $+2.42$\,dB at $\sigma_{\mathrm{m}}=2$\,dB. The
sensor-disabled controller degrades about as fast as feedback-only, so the marginal value
of sensing grows too, to $+1.80$\,dB (Fig.~\ref{fig:vsense}). Raising the alarm threshold
or admission margin keeps the controller's loss below the plain tracker's
(supplementary material).

\smallskip\noindent\textbf{Late sensing hurts; late link information favors sensing.} At Scenario~9's native
interval, a sensor one epoch late supplies the previous epoch's current-slot output. It
raises the controller's loss to $0.511$\,dB, $0.53$\,dB
($[0.00,1.12]$) below the plain tracker's but $0.25$\,dB above the sensor-disabled
controller's. At two
epochs the excess is $0.42$\,dB (Fig.~\ref{fig:vsense}). With current link observations, a late sensor
is thus worse than none. Late link information instead favors sensing. With every link-power observation one epoch stale, the controller gains
$+1.16$\,dB over the plain tracker and the marginal value of sensing is $+0.57$\,dB; with
the sensor also one epoch late, sensing still adds $+0.27$\,dB.
Applying every commitment one epoch late leaves $+0.94$\,dB over the plain tracker. If
probe outcomes arrive one epoch late, so that a probing epoch serves its pre-probe beam,
the controller keeps $+0.96$\,dB over the plain tracker and the marginal value of
sensing stays positive (Fig.~\ref{fig:vsense}). Verification can then no longer act
within the epoch, however, and always-on sensing, whose verified variants then coincide with it, beats the
controller by $0.17$\,dB
($[0.04,0.29]$).

The controller's margin over always-on sensing in Scenario~9 is therefore a same-epoch
result. Its protection at the low-innovation sites does not depend on same-epoch timing:
always-on sensing stays
$0.2$--$2.2$\,dB behind feedback-only there under every delay, while the controller stays
within $0.12$\,dB of feedback-only (supplementary material). With sensor output and probe
outcomes both one epoch late, the controller still gains $+0.58$\,dB ($[0.04,1.18]$)
over feedback-only but falls $0.20$\,dB behind the sensor-disabled controller
(Fig.~\ref{fig:vsense}). End-to-end NR timing compliance is not
established.

\section{Discussion and Conclusion}\label{sec:discussion}
\noindent\textbf{Main findings.}\label{sec:conclusion}
We asked when sensing adds useful information beyond link feedback, and when it should
be withheld. The usefulness of sensing depends jointly on the limitations of link
feedback, the loss it leaves or the probing it needs, and on the accuracy and timeliness of
the sensing information. Where feedback
already comes within $0.25$\,dB of the oracle, as in three of the four scenarios, always-on sensing is harmful. The controller instead loses at most
$0.02$\,dB to feedback-only while invoking the model on $1.0$--$6.8\%$ of frames. In
high-innovation Scenario~9 the controller's system gain is $+0.82$\,dB over the plain
feedback-only tracker and $+0.15$\,dB over a probe-limited re-sweeping reference, with fewer probes than
either. It invokes the model on $33.0\%$ of frames, against $85.2\%$ for always-on sensing. Most of the $+0.82$-dB gain corrects the setup lock loss of seven of the $28$ test passes, which either the re-sweep or the sensor removes. With both, the marginal value
of sensing at the native interval is not resolved in loss, and sensing instead
substitutes for $21\%$ of the candidate-beam probes. When feedback is stressed by noise,
staleness or longer decision intervals, an accurate sensor also reduces the same
controller's loss. Weak sensing can instead become harmful at longer decision intervals,
and with current link observations a stale sensor output is worse than none.

\looseness=-1
\smallskip\noindent\textbf{Interpretation.} One validation-selected constant
set makes these decisions without a regime label, and constants selected without
Scenario~9 keep $+0.73$\,dB there (Sec.~\ref{sec:voi_controller}). The ablations assign the roles: current sensing information supplies the
proposals, verification protects the link where the sensor is weak, and the blacklist
limits model calls; the innovation gate is nearly inert. Cross-scenario comparisons do not isolate the effects of geometry, illumination or sensing quality.

\smallskip\noindent\textbf{Scope.} The evidence covers four $60$-GHz scenarios from three
base-station sites on two streets in one dataset family, with a fixed $64$-beam codebook,
at low vehicle speeds (medians of $2.5$--$4.9$\,m/s from GPS). Only one site, where
vehicles pass close to the base station, leaves a plain tracker a shortfall at the native interval. The larger gains at longer decision intervals come from decimating
that site with cached native-cadence predictions; they indicate, rather than measure,
what slower feedback would give. Scenario~32 has only three test
passes and shares its road with Scenario~33. These results do not establish transfer
of the gains or controller constants to new sites, sensors, sensing models (here arg\,max classifiers), codebooks, or frequency
bands.

\looseness=-1
\smallskip\noindent\textbf{Limitations.} Causal replay restricts each policy's
observations but still relies on exhaustive recorded sweeps and idealized within-epoch
timing. The $64$ beams of a sweep were measured sequentially, and a fixed beam's
inter-epoch power change has a median of $1.42$\,dB, above the $0.5$-dB admission margin
(supplementary material). Delayed probe outcomes and independent measurement errors
(Sec.~\ref{sec:voi_tail}) test this only partly, since within-sweep timing errors cannot be
reconstructed. The re-sweep entered the design after a test-set analysis exposed the setup lock loss
of Sec.~\ref{sec:voi_day}, so Scenario-9 comparisons that involve it are test-informed
rather than confirmatory. Constants selected before
that analysis, without the re-sweep, already gain $+0.65$\,dB ($[0.19,1.17]$) over the plain
tracker (supplementary material). A candidate second high-innovation site, Scenario~35, failed a screening rule
fixed before any training. Its power profile is nearly flat, and feedback-only tracking leaves $0.16$--$0.22$\,dB to recover
(supplementary material). Two priorities remain: a high-innovation site with usable link
margin, and end-to-end timing measurements. The $205$-policy frontier is a measured
comparison family, not an exhaustive lower bound. It does not include learned
feedback-only trackers, so the reported crossing and probing ratio are conditional on
the tested family and its candidate-probe accounting.

\smallskip\noindent\textbf{Deployment implications.} For an illustrative mapping with one candidate-beam measurement per symbol, one probe takes roughly $9$\,$\mu$s at $120$\,kHz subcarrier spacing. The $0.87$ probes/epoch that sensing saves against the sensor-disabled
controller are therefore negligible in one link's airtime; all $3.25$ of the controller's
probes occupy $0.026\%$ (supplementary material). Per-link airtime can still understate a probe's cost, because an analog panel measures one beam
at a time for every user it serves. The number of measurement--report cycles depends on the
configured resources, reporting scheme and refinement procedure, and cannot be inferred from the
probe count. Counting inference alone, the $0.33$ model calls/epoch cost less than the probes they
replace whenever one call costs less than about $2.6$ probes in a common cost unit. Whether these costs bind was not
measured here, and the delay and decimation tests show that verification does not
guarantee a nonnegative marginal value of sensing when operating conditions change. Sensing should
therefore be evaluated against feedback-only references that include standard
refinements such as a post-setup re-sweep, and against the same policy with its sensor
disabled, with probe and invocation costs reported separately. Algorithm~\ref{alg:controller} and the supplementary material document the
controller, data splits, constant selection, and replay assumptions. The replay and analysis code is available at \href{https://github.com/jeongwankang95/sensing-aided-beam-tracking}{github.com/jeongwankang95/sensing-aided-beam-tracking}.

\section*{Acknowledgment}
The authors used Claude (Anthropic) to assist with editing the text (e.g., grammar) and with preparing figures and code. The authors reviewed and verified all such content and take full responsibility for it.

\bibliographystyle{IEEEtran}
\bibliography{library}

\begin{thebibliography}{10}
\providecommand{\url}[1]{#1}
\csname url@samestyle\endcsname
\providecommand{\newblock}{\relax}
\providecommand{\bibinfo}[2]{#2}
\providecommand{\BIBentrySTDinterwordspacing}{\spaceskip=0pt\relax}
\providecommand{\BIBentryALTinterwordstretchfactor}{4}
\providecommand{\BIBentryALTinterwordspacing}{\spaceskip=\fontdimen2\font plus
\BIBentryALTinterwordstretchfactor\fontdimen3\font minus
  \fontdimen4\font\relax}
\providecommand{\BIBforeignlanguage}[2]{{%
\expandafter\ifx\csname l@#1\endcsname\relax
\typeout{** WARNING: IEEEtran.bst: No hyphenation pattern has been}%
\typeout{** loaded for the language `#1'. Using the pattern for}%
\typeout{** the default language instead.}%
\else
\language=\csname l@#1\endcsname
\fi
#2}}
\providecommand{\BIBdecl}{\relax}
\BIBdecl

\bibitem{heath2016overview}
R.~W. Heath, N.~Gonz{\'a}lez-Prelcic, S.~Rangan, W.~Roh, and A.~M. Sayeed, ``An
  overview of signal processing techniques for millimeter wave {MIMO}
  systems,'' \emph{IEEE J. Sel. Topics Signal Process.}, vol.~10, no.~3, pp.
  436--453, Apr. 2016.

\bibitem{niu2015survey}
Y.~Niu, Y.~Li, D.~Jin, L.~Su, and A.~V. Vasilakos, ``A survey of millimeter
  wave communications ({mmWave}) for {5G}: Opportunities and challenges,''
  \emph{Wireless Netw.}, vol.~21, no.~8, pp. 2657--2676, Nov. 2015.

\bibitem{3gpp38820}
{3GPP}, ``{NR}; 7--24 {GHz} frequency range,'' TR 38.820, v16.1.0, Mar. 2021.

\bibitem{kang2024uppermidband}
S.~Kang \emph{et~al.}, ``Cellular wireless networks in the upper mid-band,''
  \emph{IEEE Open J. Commun. Soc.}, vol.~5, pp. 2058--2075, 2024.

\bibitem{demirhan2022radar}
U.~Demirhan and A.~Alkhateeb, ``Radar aided {6G} beam prediction: Deep learning
  algorithms and real-world demonstration,'' in \emph{Proc. IEEE Wireless
  Commun. Netw. Conf.}, 2022, pp. 2655--2660.

\bibitem{ma2025kd}
M.~Ma, N.~T. Nguyen, N.~Shlezinger, Y.~C. Eldar, A.~L. Swindlehurst, and
  M.~Juntti, ``Knowledge distillation for sensing-assisted long-term beam
  tracking in {mmWave} communications,'' \emph{IEEE Trans. Wireless Commun.},
  vol.~25, pp. 20\,494--20\,511, 2026.

\bibitem{charan2022challenge}
G.~Charan, U.~Demirhan, J.~Morais, A.~Behboodi, H.~Pezeshki, and A.~Alkhateeb,
  ``Multi-modal beam prediction challenge 2022: Towards generalization,'' 2022,
  arXiv:2209.07519.

\bibitem{precursor}
J.~Kang, M.~Bordallo~L\'{o}pez, N.~T. Nguyen, J.~Kim, and S.~Kim, ``Radar-only
  long-term beam prediction in {mmWave} vehicular networks: An empirical study
  of {Doppler} input and cross-modal distillation,'' 2026, arXiv:2610.20185.

\bibitem{vibe2026}
A.~Biswas, A.~Pramanik, E.~Ekici, and M.~C. Vuran, ``Look once, beam twice:
  Camera-primed real-time double-directional {mmWave} beam management for
  vehicular connectivity,'' in \emph{Proc. 22nd Annu. IEEE Int. Conf. Sens.
  Commun. Netw.}, 2026, pp. 533--542.

\bibitem{10144504}
A.~Alkhateeb \emph{et~al.}, ``{DeepSense} {6G}: A large-scale real-world
  multi-modal sensing and communication dataset,'' \emph{IEEE Commun. Mag.},
  vol.~61, no.~9, pp. 122--128, Sep. 2023.

\bibitem{atiekf2024}
M.~Mo, C.~Liu, L.~Zhao, M.~Guo, M.~Li, and Y.~Wang, ``Iterated extended
  {Kalman} filter based adaptive beam tracking for millimeter-wave systems,''
  in \emph{Proc. IEEE/CIC Int. Conf. Commun. China}, 2024, pp. 1585--1590.

\bibitem{zakeri2025drl}
A.~Zakeri, N.~T. Nguyen, A.~Alkhateeb, and M.~Juntti, ``Deep reinforcement
  learning for dynamic sensing and communications,'' in \emph{Proc. IEEE Int.
  Conf. Acoust. Speech Signal Process.}, 2026, pp. 20\,636--20\,640.

\bibitem{zakeri2025aoi}
A.~Zakeri, N.~T. Nguyen, A.~Alkhateeb, and M.~Juntti, ``{AoI}-aware machine
  learning for constrained multimodal sensing-aided communications,'' in
  \emph{Proc. IEEE Int. Conf. Commun.}, 2026, pp. 1--6.

\bibitem{probing2026adaptive}
A.~Orimogunje \emph{et~al.}, ``Sensing-assisted adaptive beam probing with
  calibrated multimodal priors and uncertainty-aware scheduling,'' \emph{IEEE
  Wireless Commun. Lett.}, vol.~15, pp. 2438--2442, 2026.

\bibitem{luo2023radar}
H.~Luo, U.~Demirhan, and A.~Alkhateeb, ``Millimeter wave {V2V} beam tracking
  using radar: Algorithms and real-world demonstration,'' in \emph{Proc. 31st
  Eur. Signal Process. Conf.}, 2023, pp. 740--744.

\bibitem{charan2022vision}
G.~Charan, T.~Osman, A.~Hredzak, N.~Thawdar, and A.~Alkhateeb,
  ``Vision-position multi-modal beam prediction using real millimeter wave
  datasets,'' in \emph{Proc. IEEE Wireless Commun. Netw. Conf.}, 2022, pp.
  2727--2731.

\bibitem{jiang2022cv}
S.~Jiang and A.~Alkhateeb, ``Computer vision aided beam tracking in a
  real-world millimeter wave deployment,'' in \emph{Proc. IEEE Globecom
  Workshops}, 2022, pp. 142--147.

\bibitem{klautau2019lidar}
A.~Klautau, N.~Gonz{\'a}lez-Prelcic, and R.~W. Heath, ``{LIDAR} data for deep
  learning-based {mmWave} beam-selection,'' \emph{IEEE Wireless Commun. Lett.},
  vol.~8, no.~3, pp. 909--912, Jun. 2019.

\bibitem{jiang2023lidar}
S.~Jiang, G.~Charan, and A.~Alkhateeb, ``{LiDAR} aided future beam prediction
  in real-world millimeter wave {V2I} communications,'' \emph{IEEE Wireless
  Commun. Lett.}, vol.~12, no.~2, pp. 212--216, Feb. 2023.

\bibitem{Tian2023MultimodalTF}
Y.~Tian, Q.~Zhao, Z.~E.~A. Kherroubi, F.~Boukhalfa, K.~Wu, and F.~Bader,
  ``Multimodal transformers for wireless communications: A case study in beam
  prediction,'' \emph{ITU J. Future Evolving Technol.}, vol.~4, no.~3, pp.
  461--471, Sep. 2023.

\bibitem{shi2024multimodal}
B.~Shi, M.~Li, M.-M. Zhao, M.~Lei, and L.~Li, ``Multimodal deep learning
  empowered millimeter-wave beam prediction,'' in \emph{Proc. IEEE 99th Veh.
  Technol. Conf.}, 2024, pp. 1--6.

\bibitem{zhu2025crossmodal}
Q.~Zhu, Y.~Wang, W.~Li, H.~Huang, and G.~Gui, ``Advancing multi-modal beam
  prediction with cross-modal feature enhancement and dynamic fusion
  mechanism,'' \emph{IEEE Trans. Commun.}, vol.~73, no.~9, pp. 7931--7940, Sep.
  2025.

\bibitem{cheng2022isac}
X.~Cheng, D.~Duan, S.~Gao, and L.~Yang, ``Integrated sensing and communications
  ({ISAC}) for vehicular communication networks ({VCN}),'' \emph{IEEE Internet
  Things J.}, vol.~9, no.~23, pp. 23\,441--23\,451, Dec.~1, 2022.

\bibitem{yi2024beam}
Y.~Wang, Z.~Wei, and Z.~Feng, ``Beam training and tracking in {mmWave}
  communication: A survey,'' \emph{China Commun.}, vol.~21, no.~6, pp. 1--22,
  Jun. 2024.

\bibitem{liu2020radar}
F.~Liu, W.~Yuan, C.~Masouros, and J.~Yuan, ``Radar-assisted predictive
  beamforming for vehicular links: Communication served by sensing,''
  \emph{IEEE Trans. Wireless Commun.}, vol.~19, no.~11, pp. 7704--7719, Nov.
  2020.

\bibitem{yuan2021bayesian}
W.~Yuan, F.~Liu, C.~Masouros, J.~Yuan, D.~W.~K. Ng, and
  N.~Gonz{\'a}lez-Prelcic, ``Bayesian predictive beamforming for vehicular
  networks: A low-overhead joint radar-communication approach,'' \emph{IEEE
  Trans. Wireless Commun.}, vol.~20, no.~3, pp. 1442--1456, Mar. 2021.

\bibitem{park2025ckd}
Y.~M. Park, S.~S. Hassan, W.~Saad, and C.~S. Hong, ``Cross-modal knowledge
  distillation for efficient radar-only beam prediction in {mmWave}
  communications,'' in \emph{Proc. IEEE 26th Int. Workshop Signal Process.
  Artif. Intell. Wireless Commun.}, 2025, pp. 1--5.

\bibitem{shang2025multimodal}
C.~Shang, D.~T. Hoang, and J.~Yu, ``Multi-modal beamforming with model
  compression and modality generation for {V2X} networks,'' \emph{IEEE Trans.
  Mobile Comput.}, vol.~25, no.~11, pp. 20\,492--20\,506, Nov. 2026.

\bibitem{wang2025lowlight}
H.~Wang, B.~Ou, X.~Xie, and Y.~Wang, ``Vision-aided {mmWave} beam and blockage
  prediction in low-light environment,'' \emph{IEEE Wireless Commun.\ Lett.},
  vol.~14, no.~3, pp. 791--795, Mar. 2025.

\bibitem{yang2023env}
Y.~Yang, F.~Gao, X.~Tao, G.~Liu, and C.~Pan, ``Environment semantics aided
  wireless communications: A case study of {mmWave} beam prediction and
  blockage prediction,'' \emph{IEEE J. Sel. Areas Commun.}, vol.~41, no.~7, pp.
  2025--2040, Jul. 2023.

\bibitem{imran2024env}
S.~Imran, G.~Charan, and A.~Alkhateeb, ``Environment semantic communication:
  Enabling distributed sensing aided networks,'' \emph{IEEE Open J. Commun.
  Soc.}, vol.~5, pp. 7767--7786, 2024.

\bibitem{zakeri2025constrained}
A.~Zakeri, N.~T. Nguyen, A.~Alkhateeb, and M.~Juntti, ``Constrained multimodal
  sensing-aided communications: A dynamic beamforming design,'' in \emph{Proc.
  IEEE Global Commun. Conf.}, 2025, pp. 4772--4776.

\bibitem{commrad2024}
I.~K. Jain, S.~MM, and D.~Bharadia, ``{CommRad}: Context-aware sensing-driven
  millimeter-wave networks,'' in \emph{Proc. 22nd ACM Conf. Embedded Netw.
  Sensor Syst.}, 2024, pp. 633--646.

\bibitem{hussain2021dual}
M.~Hussain and N.~Michelusi, ``Learning and adaptation for millimeter-wave beam
  tracking and training: A dual timescale variational framework,'' \emph{IEEE
  J. Sel. Areas Commun.}, vol.~40, no.~1, pp. 37--53, Jan. 2022.

\bibitem{thz2022event}
Y.~Karacora, C.~Chaccour, A.~Sezgin, and W.~Saad, ``Event-based beam tracking
  with dynamic beamwidth adaptation in terahertz ({THz}) communications,''
  \emph{IEEE Trans.\ Commun.}, vol.~71, no.~10, pp. 6195--6210, Oct. 2023.

\bibitem{scirep2024beam}
L.~Marenco, L.~E. Hupalo, N.~F. Andrade, and F.~A.~P. de~Figueiredo,
  ``Machine-learning-aided method for optimizing beam selection and update
  period in {5G} networks and beyond,'' \emph{Sci.\ Rep.}, vol.~14, no.~1, Aug.
  2024, {Art.} no. 20103.

\bibitem{v2i2024rl}
J.~Lee, H.~Kim, and J.~So, ``Reinforcement learning-based joint beamwidth and
  beam alignment interval optimization in {V2I} communications,''
  \emph{Sensors}, vol.~24, no.~3, Feb. 2024, {Art.} no. 837.

\bibitem{ayan2019aoi}
O.~Ayan, M.~Vilgelm, M.~Kl{\"u}gel, S.~Hirche, and W.~Kellerer,
  ``Age-of-information vs. value-of-information scheduling for cellular
  networked control systems,'' in \emph{Proc. 10th ACM/IEEE Int. Conf.
  Cyber-Phys. Syst.}, 2019, pp. 109--117.

\bibitem{isacfc2026}
T.~Soleymani, M.~Assaad, and J.~S. Baras, ``Feedback control via integrated
  sensing and communication: Uncertainty optimisation,'' 2026,
  arXiv:2601.22912.

\bibitem{va2016beam}
V.~Va, H.~Vikalo, and R.~W. Heath, ``Beam tracking for mobile millimeter wave
  communication systems,'' in \emph{Proc. IEEE Global Conf. Signal Inf.
  Process.}, 2016, pp. 743--747.

\bibitem{3gpp38843}
{3GPP}, ``Study on artificial intelligence ({AI})/machine learning ({ML}) for
  {NR} air interface,'' TR 38.843, v18.0.0, Dec. 2023.

\bibitem{giordani2019tutorial}
M.~Giordani, M.~Polese, A.~Roy, D.~Castor, and M.~Zorzi, ``A tutorial on beam
  management for {3GPP} {NR} at {mmWave} frequencies,'' \emph{IEEE Commun.
  Surveys Tuts.}, vol.~21, no.~1, pp. 173--196, 1st Quart. 2019.

\bibitem{va2017impact}
V.~Va, J.~Choi, and R.~W. Heath, ``The impact of beamwidth on temporal channel
  variation in vehicular channels and its implications,'' \emph{IEEE Trans.
  Veh. Technol.}, vol.~66, no.~6, pp. 5014--5029, Jun. 2017.

\end{thebibliography}
\end{document}